\documentclass[trackchanges]{aastex7}%linenumbers,
\usepackage{multirow}

\usepackage[usenames,table,dvipsnames]{xcolor}
\definecolor{blue}{RGB}{66, 153, 233}
\definecolor{red}{RGB}{255, 0, 0}
\definecolor{purple}{RGB}{255, 0, 255}
\newcount\Comments  % 0 suppresses notes to selves in text
\newcommand{\kibitz}[2]{\ifnum\Comments=1\textcolor{#1}{#2}\fi}

\begin{document}

\title{Detectability of Polarized $\gamma$-Ray Emission from Blazar Flares with COSI}

\author[orcid=0009-0003-4039-9685]{Garrett A. Latiolais}
\affiliation{Department of Physics and Astronomy, Louisiana State University, LA, USA}
\email{}

\author[orcid=0000-0002-4241-5875]{Jorge Otero-Santos} 
\affiliation{Istituto Nazionale di Fisica Nucleare, Sezione di Padova, 35131 Padova, Italy}
\email{}

\author[orcid=0000-0002-6548-5622]{Michela Negro$^*$}
\affiliation{Department of Physics and Astronomy, Louisiana State University, LA, USA}
\email[show]{michelanegro@lsu.edu}

\author[orcid=0000-0002-8472-3649]{Lea Marcotulli}
\affiliation{Deutsches Elektronen-Synchrotron DESY, Platanenallee 6, 15738 Zeuthen, Germany}
\affiliation{Department of Physics and Astronomy, Clemson University, Kinard Lab of Physics, Clemson, SC 29634-0978, USA}
\email{}

\author[orcid=0009-0005-0912-8304]{Mohammad Ali Boroumand}
\affiliation{Department of Physics and Astronomy, Louisiana State University, LA, USA}
\email{}

\author[0000-0002-2664-8804]{Savitri Gallego}
\affiliation{Institut für Physik \& Exzellenzcluster PRISMA+, Johannes Gutenberg-Universität Mainz, 55099 Mainz, Germany}
\email{} 

\author[0000-0002-6774-3111]{Christopher M. Karwin}
\affiliation{Department of Physics and Astronomy, Clemson University, Clemson, SC 29634, USA}
\email{}

\author[orcid=0000-0002-2471-8696]{Israel Martinez-Castellanos}
\affiliation{Department of Astronomy, University of Maryland, College Park, MD 20742, USA}
\affiliation{NASA Goddard Space Flight Center, 8800 Greenbelt Road, Greenbelt, MD 20771, USA}
\email{}

\author[orcid=0000-0001-9201-4706]{Daniel Kocevski}
\affiliation{ST12 Astrophysics Branch, NASA Marshall Space Flight Center, Huntsville, AL 35812, USA}
\email{}

\author[orcid=0000-0001-7774-5308]{Marco Ajello}
\affiliation{Department of Physics and Astronomy, Clemson University, Clemson, SC 29634, USA}
\email{}

\author[0009-0007-1918-577X]{Sara Capecchiacci}
\affiliation{Institute of Astrophysics, Foundation for Research and Technology - Hellas, Voutes, 70013 Heraklion, Greece}
\affiliation{Department of Physics, University of Crete, 70013, Heraklion, Greece}
\email{}  

\author[0000-0001-9200-4006]{Ioannis Liodakis}
\affiliation{Institute of Astrophysics, Foundation for Research and Technology - Hellas, Voutes, 70013 Heraklion, Greece}
\email{} %liodakis@ia.forth.gr
  
%Other COSI Members in alphabetical order below

\author[orcid=0000-0003-1515-9500]{Srinadh R. Bhavanam}
\affiliation{Department of Physics and Astronomy, Clemson University, Clemson, SC 29634, USA}
\email{}  

\author[orcid=0000-0001-9567-4224]{Steven E. Boggs}
\affiliation{Department of Astronomy \& Astrophysics, University of California, San Diego, 9500 Gilman Drive, La Jolla, CA 92093-0424, USA}
\email{}

\author[0000-0002-8028-0991]{Dieter H. Hartmann}
\affiliation{Department of Physics and Astronomy, Clemson University, Clemson, SC 29634, USA}
\email{}

\author[orcid=0000-0001-6677-914X]{Carolyn A. Kierans}
\affiliation{NASA Goddard Space Flight Center, 8800 Greenbelt Road, Greenbelt, MD 20771, USA}
\email{}

\author[orcid=0000-0002-9854-1432]{Tiffany R. Lewis}
\affiliation{Department of Physics, Earth, Planetary \& Space Sciences Institute, Michigan Technological University, Houghton, MI 49931, USA}
\email{}

\author[orcid=0000-0001-6181-839X]{Alberto Sciaccaluga}
\affiliation{INAF -- Osservatorio Astronomico di Brera, Via E. Bianchi 46, I-23807 Merate, Italy}
\email{}

\author[orcid=0000-0001-5506-9855]{John A. Tomsick}
\affiliation{Space Sciences Laboratory, 7 Gauss Way, University of California, Berkeley, CA 94720-7450, USA}
\email{}

\author[0000-0001-9826-1759]{Haocheng Zhang}
\affiliation{University of Maryland Baltimore County Baltimore, MD 21250, USA}
\affiliation{NASA Goddard Space Flight Center Greenbelt, MD 20771, USA}
\email{}

\author[orcid=0000-0001-5506-9855]{Andreas Zoglauer}
\affiliation{Space Sciences Laboratory, 7 Gauss Way, University of California, Berkeley, CA 94720-7450, USA}
\email{}

% \author{Other COSI, alphabetical order}
% \affiliation{Institution}
%  \email{}

% \tableofcontents

\begin{abstract}
We investigate the detectability of polarized $\gamma$-ray emission from blazar flares with the Compton Spectrometer and Imager (COSI). Using 17 years of Fermi Large Area Telescope observations, we analyze light curves for {1413} blazars and identify {a maximum of 787 sources with} flaring episodes through Bayesian block analysis. For each flare, we estimate the minimum detectable polarization ($\mathrm{MDP}_{99\%}$) in the COSI energy band {(0.2-5 MeV)} using instrument response functions under a range of spectral assumptions and background conditions.
Under baseline background levels (1 {counts}/s), and assuming that blazar flare statistics in the MeV band are comparable to those observed at GeV energies, we find that COSI can realistically detect polarization in {up to $\sim$6} flares with $\mathrm{MDP}_{99\%}<50\%$ over its two-year prime mission {depending on different spectral and flare identification assumptions}, with {only a few most powerful ones} reaching $\mathrm{MDP}_{99\%}<20\%$. {These expectations are shown to improve when shorter intervals around bright peaks within long flares are considered.} We provide a ranked list of the most promising targets, finding that flat-spectrum radio quasars dominate the population of polarization-detectable events.
Through its continuous all-sky monitoring in the largely unexplored MeV band, COSI will open a new observational window on blazar variability and deliver the first direct measurements of MeV polarization, offering unique insights into jet geometry and high-energy emission processes.

\end{abstract}


\section{Introduction}\label{sec:introduction}

\setcounter{footnote}{0}

Active Galactic Nuclei (AGNs) are found in the core of some galaxies, in which a supermassive black hole is actively accreting gas and sometimes can produce collimated jets of material and radiation. The jet can accelerate particles to relativistic regimes \citep[see e.g.][]{blandford1979}, which produces high-energy radiation up to TeV energies. AGNs whose jets point towards the line-of-sight within $\sim 5^\circ$ are classified as blazars, which are further subclassified (depending on their optical spectrum) into Flat-Spectrum Radio Quasars (FSRQs) if strong broad ($|EW|>5$~\AA) optical emission lines are observed, and BL Lacertae (BL Lac) objects if {weak} to no optical lines are observed \citep{urry1995}. 

Blazar emission extends from radio to $\gamma$ rays, covering the entire electromagnetic spectrum. This multi-wavelength emission, characterized by its remarkable variability on all possible timescales \citep[see e.g.][]{ryan2019,raiteri2023,otero2024}, shows a two-hump shape in its typical spectral energy distribution (SED) representation. The low-energy emission is well understood and explained as synchrotron radiation of relativistic electrons moving under the influence of the magnetic field {in the} jets \citep{koenigl1981}. However, the nature of the high-energy emission still remains debated, being one of the most pressing questions on blazar, high-energy and astroparticle physics. The most common interpretation --- the leptonic scenario --- assumes that the high-energy emission is produced via inverse Compton (IC) scattering of low-energy photons with the relativistic electrons. This can happen with low-energy synchrotron photons \citep[Synchrotron Self-Compton scattering or SSC, see][]{maraschi1992} or with external low-energy photons from outside the jet \citep[External Compton scattering or EC, see][]{dermer1993}. Alternatively, hadronic interactions have also been proposed as a viable way to explain their high-energy component \citep{Mannheim1993}. This scenario has become especially relevant with the possible association of blazars with the emission of extragalactic astrophysical neutrinos \citep{aartsen2018}.

High-energy and multi-wavelength polarimetry has emerged as a powerful new observable for probing the high-energy emission of astrophysical sources. With the recent launch of the Imaging X-ray Polarimetry Explorer (IXPE) \citep{weisskopf2022}, operating in the soft X-ray band, X-ray polarization measurements have provided unprecedented insight into the synchrotron (low-energy) hump of blazar spectral energy distributions. These observations have yielded strong constraints on the magnetic-field geometry and particle-acceleration mechanisms in relativistic jets, notably disfavoring simple one-zone emission models \citep[see, e.g.,][]{middei2023, peirson2023, agudo2025}. However, despite these advances, the dominant composition of blazar jets—whether primarily leptonic or hadronic—remains an open question. At higher energies, the Compton Spectrometer and Imager (COSI) mission \citep{tomsick2024}, scheduled for launch in 2027, will extend polarimetric capabilities into the $\gamma$-ray band, covering the 0.2–5 MeV energy range. Polarization measurements of blazar jets at $\gamma$-ray energies provide a powerful diagnostic for distinguishing between leptonic and hadronic jet compositions. Early studies demonstrated that these scenarios can lead to markedly different polarization signatures under idealized magnetic-field configurations \citep{chang2013, zhang2013}, while more recent work incorporating more realistic magnetic-field geometries predicts reduced polarization fractions overall, with leptonic scenarios yielding particularly low levels of polarization \citep{2019ApJ...876..109Z, 2024ApJ...967...93Z}. As a result, any significant detection of $\gamma$-ray polarization would strongly favor a hadronic origin. As such, COSI’s $\gamma$-ray polarimetry has the potential to represent a turning point in our understanding of the most powerful particle-acceleration phenomena in the Universe.

Over the past 17 years, the Large Area Telescope (LAT) on board the \textit{Fermi} Gamma-ray Space Telescope has played a pivotal role in advancing our understanding of the $\gamma$-ray Universe. Since its launch in 2008, the LAT has observed photons over an energy range spanning from $\sim$20 MeV to beyond 300 GeV \citep{atwood2009}. Thanks to its sensitivity and continuous all-sky monitoring, more than 5,700 sources have been detected and characterized in the high-energy $\gamma$-ray band, over 3,000 of which are blazars \citep{ballet2023}. This extensive dataset has enabled detailed studies of blazar $\gamma$-ray emission and variability across a wide range of timescales.
One notable data product of this continuous monitoring is the \textit{Fermi}-LAT Light Curve Repository \citep[LCR; see][]{abdollahi2023}, which collects and continuously updates flux measurements for variable sources, providing long-term light curves for more than 1,500 $\gamma$-ray sources included in the \textit{Fermi}-LAT Fourth Source Catalog Data Release 2 \citep[4FGL-DR2; see][]{abdollahi2020, ballet2020}. In this work, we make extensive use of this database.

In view of the upcoming launch of COSI and its role as a {daily} all-sky monitor, we aim to assess the detectability of polarized $\gamma$-ray emission from blazars, with a particular focus on flaring states. Owing to the strong flux variability of blazars and the expectation of polarization-angle variations during flares, we concentrate on relatively short, bright events (i.e., with durations shorter than $\sim$8 weeks). Our objective is to provide quantitative expectations for the number and characteristics of flares for which COSI can achieve meaningful polarization measurements.

The paper is structured as follows: in Sect.~\ref{sec:analysis} we describe our data sample, the process for identifying blazar flares, the extrapolation to the expected flux in the COSI band, and the computation of the minimum detectable polarization. In Sect.~\ref{sec:results} we present and discuss the results. In Sect.~\ref{sec:conclusion} we summarize our work and present our conclusions.

\section{Data Analysis}
\label{sec:analysis}
\subsection{Blazar $\gamma$-Ray Flare Definition} \label{sec:flare_extraction}
% - Light Curve Repository\\ 
We analyze 17 years of blazar light curves provided by the LCR. The details of the automated data analysis are provided in \cite{abdollahi2023}; here we focus on the data post-processing adopted to clean the light curves, following the caveats noted by the LAT Collaboration\footnote{\url{https://fermi.gsfc.nasa.gov/ssc/data/access/lat/LightCurveRepository/about.html}}.
For this study, we restricted our analysis to the weekly-cadence light curves, which offers the optimal balance between statistical significance and temporal resolution. The weekly binning provides sufficient photon statistics to probe COSI’s polarization capability while still resolving flares shorter than a month that may remain bright enough for polarization measurements.  We consider the light curves derived from the LCR analysis with the spectral indices left free to vary in the maximum likelihood fit \citep[see][for details]{abdollahi2023}. We filter out time bins with null or negative test statistics. In total we have 1413 blazar sources in our sample, of which  572 are FSRQs, and  477 are BL Lac objects, and 364 are BCUs \citep[blazars of uncertain type as reported in the 4FGL catalog, see][]{ballet2023}.

\begin{figure}
    \centering
    \includegraphics[width=1\linewidth]{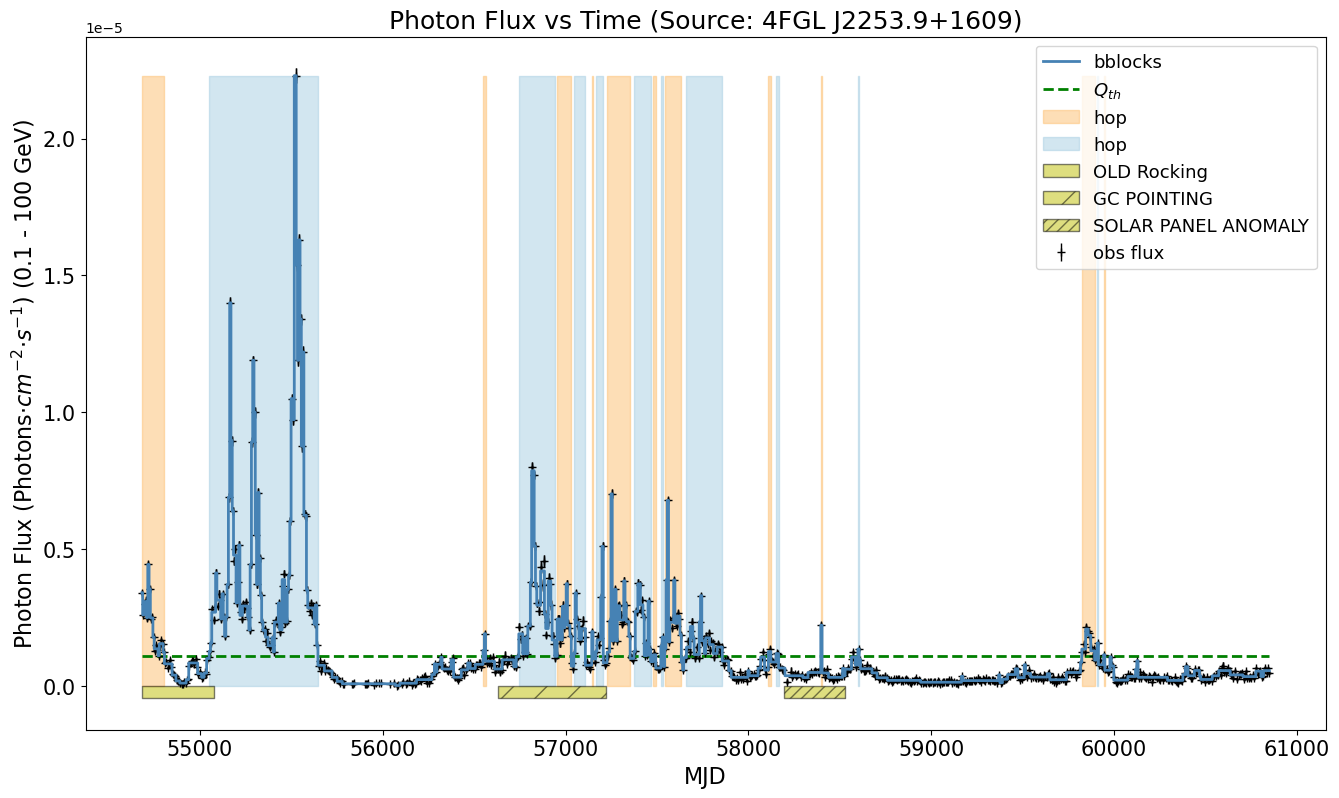}
\caption{The photon flux light curve of LCR source 4FGL J2253.9+1609, within \textit{Fermi}'s energy band with $\eta=0.3$. Alternating orange and blue blocks represent distinct identified flaring periods. The solid blue line is the {BB analysis} of the source. The dashed horizontal green line is the calculated quiescent {threshold} of the source, $Q_{th}$. Below the light curve, yellow blocks mark the periods of time where \textit{Fermi}-LAT was operating in a non-standard survey mode.}
    \label{fig:lighcurve}
\end{figure}

We proceed to the flare extraction through a Bayesian block (BB) analysis. We utilized the python package from \cite{wagner2022}\footnote{\url{https://github.com/swagner-astro/lightcurves}} to perform a BB analysis \citep[according to][]{Scargle2013} and to define and characterize flares as groups of blocks \citep[following][]{Mayer2019}. We devised the following iterative procedure to identify flares based on the BB-binned light curves. 
First, we perform the BB analysis assuming a p-value of 0.05 adopting a flux threshold for each source defined as
\begin{equation}
    \rm{F}^{\it{i}}_{\rm{th}} = \rm{F}^{\it{i}}_{\rm{min}} + \eta \cdot \Delta F^{\it{i}},
    \label{eq:flux_threshold}
\end{equation}
where $\rm{F}^{\it{i}}_{\rm{min}}$ is the minimum flux of a source over the entire light curve and $\eta$ is a fraction of the maximal variability amplitude, that is, the difference between the maximum and minimum flux points of the light curve,
\begin{equation}
    \Delta \rm F^{\it{i}} = \rm{F}^{\it{i}}_{\rm{max}}-\rm{F}^{\it{i}}_{\rm{min}}.
    \label{eq:delta_f}
\end{equation}
The fraction $\eta$ is chosen arbitrarily; we explore values of 0.1, 0.3, and 0.5, and present and discuss the corresponding results in Sect.~\ref{sec:results}. We then apply the \texttt{hop} algorithm in \texttt{baseline} mode, as described in \citet{wagner2022}, to identify quiescent periods, defined as BB intervals with fluxes below the threshold specified in Eq.~(\ref{eq:flux_threshold}). For each light curve, we compute the average flux across all quiescent intervals to determine the final quiescent flux {threshold}, $Q_{\rm th}$. This value is subsequently adopted as a refined threshold to re-run the \texttt{hop} algorithm on the BB-binned light curve, enabling the identification of flaring periods for each source.
%By averaging the flux level in quiescence we define the new threshold $F^i_{th, q}$, and we use it to iterate again the \texttt{hop} algorithm on the BB-binned light curve and identify the flaring periods for our sources. 
As an example, in Fig.~\ref{fig:lighcurve}, we show the weekly binned light curve for 4FGL~J2253.9+1609, where in blue we report the result of BB analysis, and the colored bands (alternating between blue and orange for visualization purposes) mark different identified flares. Underneath the light curve, yellow blocks mark the periods of time where \textit{Fermi}-LAT was operating in a non-standard survey mode. These refer to the original rocking angle profile, galactic center monitoring, and the period of time between the solar panel anomaly and the start of the new (and current) survey mode \citep{abdollahi2023}. This mostly served as a way to monitor whether anomalous data points would correspond to the times of transition between different observation modes, in which case we removed the data points and repeated the whole procedure. An illustration of using $F_{th}$ and  $Q_{th}$ is provided in Fig. \ref{fig:threshold_illust} of the Appendix.

{for the case of $\eta=0.3$, our BB analysis identifies a total of 787 sources that contain flares in their light curves.} For each source, each flare is tagged with its source's name, start time, and duration of the flare (sum of the BB bin widths that define the flare). {After the identification of these flares,} we can now proceed to determine the expected statistics for each of them in the COSI band, as described in the next section. {We note that there is no unique prescription for defining flares in blazar light curves, and different approaches have been adopted in the literature \citep[see e.g.][]{Nalewajko2013}. As a result, alternative flare identification methods could, in principle, lead to a different flare sample. Nevertheless, the definition adopted here yields a sample that is both sufficiently large and diverse to be representative of the flaring behavior of gamma-ray blazars. In particular, as discussed in Appendix \ref{app:2}, we validate our flare identification pipeline by recovering a duty cycle consistent with previously reported values.}

\subsection{Extrapolation to COSI Energy Band {and Event Rate Estimates}} \label{sec:extrapolation}

COSI operates in a lower energy band (0.2-5 MeV) than \textit{Fermi}-LAT (20 MeV to $>$300000 MeV). In order to assess the detectability of polarized emission from blazar flares by COSI we need to extrapolate the observations performed by \textit{Fermi}-LAT to the $0.2-5$~MeV band. In particular, we are interested in calculating the total number of photons that COSI could detect for each flare identified with the methodology described in Sect.~\ref{sec:flare_extraction}. The flare photon statistics will allow us to estimate the minimum detectable polarization at 99\% confidence level ($\mathrm{MDP}_{99\%}$, see Sect.~\ref{sec2.3}), which is standard figure of merit chosen to evaluate polarization capabilities of an instrument. %{I defined here the MDP for now but this can be moved if we want to define it before}

\begin{figure}
    \centering
        \includegraphics[width=0.5\columnwidth]{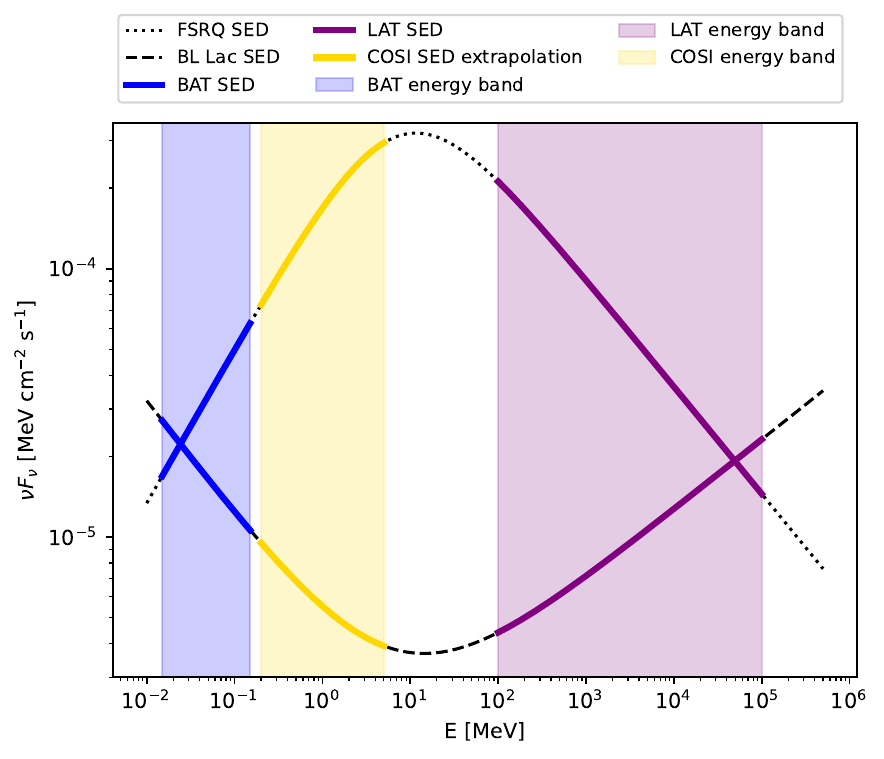}
    \caption{Example of a BAT-LAT extrapolated spectrum to the COSI 0.2-5 MeV $\gamma$-ray band according to Eqs.~(\ref{eq:BAT_LAT_spectrum1}) and (\ref{eq:BAT_LAT_spectrum2}) for a BL Lac object (dashed black line) and an FSRQ (dotted black line). The energy ranges of BAT, COSI and \textit{Fermi}-LAT are highlighted with the blue, yellow and purple shaded regions and solid lines, respectively.}
    \label{fig:extrapolation}
\end{figure}

A subset of 65 sources in our sample have been identified by \cite{tsuji2021} as both LAT sources \citep[4FGL,][]{abdollahi2020} and \textit{Swift}-BAT (15-150 keV) sources \citep[105-month BAT catalog,][]{oh2018}. For these sources, it is possible to confidently extrapolate the spectrum in the COSI energy range by jointly fitting BAT and LAT data that bracket the MeV band. We utilize the spectral characterization performed by \cite{marcotulli2022}, which describes the spectrum of these blazars with one of the following differential photon flux spectral shapes depending on the blazar's type:
\begin{equation}
\frac{dN}{dE} = K \left[ \left( \frac{E}{E_b}\right)^{\gamma_1} + \left( \frac{E}{E_b}\right)^{\gamma_2} \right]^{-1},
\label{eq:BAT_LAT_spectrum1}
\end{equation}
\begin{equation}
\frac{dN}{dE} = K \left[ \left( \frac{E}{E_b}\right)^{\gamma_1} + \left( \frac{E}{E_b}\right)^{\gamma_2} \right],
\label{eq:BAT_LAT_spectrum2}
\end{equation}
where $K$ is the normalization constant in units of photons/cm$^{2}$/s/MeV, $E_b$ is the break energy and $\gamma_1$ and $\gamma_2$ are the spectral indices below and above the break, respectively {\citep[see][]{marcotulli2022}.} 
Equation (\ref{eq:BAT_LAT_spectrum1}) used for the sources shows a hard/rising X-ray photon index ($\Gamma_{\rm X}<=2$) and a soft/falling $\gamma$-ray one  ($\Gamma_{\rm X}>=2$), indicative of the fact that the high-energy IC emission peaks in the MeV band (typical for FSRQs and low-synchrotron peaked BL~Lacs). Equation (\ref{eq:BAT_LAT_spectrum2}) instead describes sources with soft/falling X-ray spectra ($\Gamma_{\rm X}>=2$) and hard/rising $\gamma$-ray ones. The change in slopes marks the transition between the synchrotron component of the SED and the IC one (typical of high-synchrotron peaked BL Lacs), which would be detectable in the MeV. 
These two spectral shapes are used to calculate the spectrum in the COSI and LAT bands. In Fig.~\ref{fig:extrapolation} we show two examples of SED that follow the two spectral models in Eqs. (\ref{eq:BAT_LAT_spectrum1}) and (\ref{eq:BAT_LAT_spectrum2}). The blue, purple and yellow bands highlight the BAT, LAT and COSI energy ranges, respectively. 

For the remaining 1348 sources in our sample, in lack of lower-energy BAT data,  we assume the average log-parabola spectral shape reported in the 4FGL-DR3 catalog and extrapolate it to the COSI band. We note that for some sources, the 4FGL report a preferred spectral shape modeled as a power law or power law with exponential cut-off. However, to be conservative in the extrapolation, we adopt the log-parabola model also in such cases, as a power-law spectrum extrapolation would results in a larger flux overestimation in the COSI band.
By convolving the differential photon spectra with COSI’s {sky-fraction--weighted effective area, $\mathcal{E}(E)_{\rm COSI}$ (computed as described in Appendix~\ref{app:aeff})} and integrating over the $0.2$--$5$~MeV energy range, we estimate the expected photon rate in the COSI band for each source. We apply the same procedure to the LAT band, using the corresponding LAT {$\mathcal{E}(E)_{\rm LAT}$}, and then compute the ratio $f_{\rm rate} = N_{\rm COSI}/N_{\rm LAT}$, which defines the scaling factor applied to the LAT light curves for each source in our sample.
In this calculation, we assume a constant spectral index over the entire light curve of each source, adopting the average spectral index as our baseline case. We also explore alternative scenarios in which the spectral index is varied, as discussed in Sect.~\ref{sec:results}.

To estimate the expected COSI counts during a flare, we multiply the \textit{Fermi}-LAT photon flux values in the flare light curve by the spectrum-weighted average LAT exposure\footnote{Since the LCR light curves report photon fluxes already integrated over the $0.1$–$100$~GeV energy range, we adopt an average LAT exposure weighted by the spectral shape of each source, as described in Appendix~\ref{app:aeff}.}. We then apply the LAT-to-COSI scaling factor $f_{\rm rate}$ to account for the different energy bands. Finally, we integrate over the flare duration to obtain the total expected counts.

In general, in the X-ray band the spectral index is expected to deviate from its time-averaged value during flaring episodes \citep{giommi2021}. For this reason, we repeat the procedure described above for the 65 sources with available X-ray spectral information, assuming both a hardening and a softening of the X-ray spectrum prior to computing $f_{\rm rate}$. Specifically, we vary the spectral index $\gamma_1$ (below the break energy $E_b$) in Eqs.~(\ref{eq:BAT_LAT_spectrum1}) and (\ref{eq:BAT_LAT_spectrum2}), following systematic X-ray spectral studies of large blazar samples that exhibit both softer-when-brighter and harder-when-brighter trends \citep[see, e.g.,][]{giommi2021}.
We modify $\gamma_1$ in steps of 0.025, spanning a range of $\pm0.5$ around the baseline value. For each case, we compute a new set of COSI-detectable flares and evaluate their polarization detectability, as described in the following section. In Sect.~\ref{sec:results}, however, we report only the most favorable (“best-case”) scenario, which corresponds to a softening of the spectrum by $+0.5$.

\subsection{COSI's Minimum Detectable Polarization for flares}\label{sec2.3}

The minimum detectable polarization ($\mathrm{MDP}_{99\%}$) is the smallest polarization fraction that, with 99\% confidence, an instrument would measure as being distinguishable from an unpolarized source. It depends on the source and background counts collected during the observation as follows {\citep[see e.g.][]{Weisskopf2010}}:
\begin{equation}
\mathrm{MDP}_{99\%}=\frac{4.29}{\mu N_S}\sqrt{N_S + N_B},
\label{eq:MDP}
\end{equation}
where $N_S$ and $N_B$ denote the number of detected source and background photons, respectively, and $\mu$ is the average modulation factor, defined as the instrument response to 100\% polarized radiation over the considered energy range. {The background photons are assumed to be unpolarized.} For the current COSI mission design, the average modulation factor has been determined through Monte Carlo simulations based on the full instrument mass model and {monochromatic} 100\% polarized beams at different energies. Because the effective area of COSI dominates the calculation of the modulation response, we adopt an effective-area–averaged constant value of $\bar{\mu} = 0.3$ for all flares in our sample. Additional details on the estimation of the modulation factor are provided in Appendix~\ref{app:1}.

The estimated background count rate for COSI has been obtained from the simulated instrumental and astrophysical background available in the latest COSI Data Challenge \citep[DC3,][]{zoglauer2021,martinez-castellanos2023,DC3_zenodo}\footnote{https://github.com/cositools/cosi-data-challenges}. Astrophysical background contributions come from atmospheric albedo emission, the extragalactic $\gamma$-ray background, and diffuse Galactic emission. Instrumental backgrounds, on the other hand, are produced by cosmic rays interacting with the detector. These consist of a prompt component, generated when cosmic-ray particles directly trigger the instrument, and a delayed component, originating from radioactive decay of materials activated by irradiation. The dominant contributors are primary protons, $\alpha$-particles, electrons, and positrons, as well as atmospheric neutrons and secondary protons, electrons, and positrons.
All of these are accounted for in the DC3 background model, as detailed in the documentation online, and yield to a total count rate in the whole band of $\sim$20 {counts}/s\footnote{{Here, “counts” refers to reconstructed photon-like detections and should not be confused with detector-level “hits” produced by a single particle in one or multiple detectors.}} \citep[abbrev. ct/s, ][]{gallego2025}. {However, when we make use of COSI’s imaging capabilities, only $\sim$5\% of these background events are consistent with a particular point source. Therefore, after applying spatial cuts to isolate the photons consistent with the direction of an AGN \citep{Zoglauer2006}, the baseline count rate is 1 ct/s.}
%All of these are accounted for in the DC3 background model, as detailed in the documentation online, and yield to a total count rate in the whole band of $\sim$ 20 ph/s \citep{gallego2025}. This corresponds to the full sky background count rate, which is expected to be reduced by a factor of $\sim20$ applying appropriate spatial cuts to isolate the photons consistent with the sky location of a point-like source of interest {\citep{Zoglauer2006}}. Hence, our baseline count rate is 1~ph/s. 
The total background counts for each flare can be easily determined by multiplying the expected rate by the flare duration.

\section{Results and Discussion}\label{sec:results}

%In our results 
{Initially,} we focus only on flares with duration below 8 weeks, as polarization observations in other bands show highly variable polarization angles and polarization degree on long timescales. For example, optical observations from RoboPol show that large rotations of the polarization angle within several days are common, especially among low synchrotron peaked blazars \citep{2018MNRAS.474.1296B}. While time scales can vary between different blazar subclasses, we have found that the typical time periods of polarization angle stability range from two to four weeks on average (Capecchiacci et al., 2026, in preparation). Additionally, theoretical studies based on three-dimensional simulations that self-consistently couple particle evolution with radiation transport show that hadronic polarization signatures evolve on longer timescales than their leptonic counterparts, consistent with the timescales considered in this work \citep{2016ApJ...829...69Z}. 
%{Therefore, flares with a duration longer than 8 weeks have been excluded from the $\mathrm{MDP}_{99\%}$ estimation.}

We consider our baseline background rate to be 1 {ct}/s. This background rate reduction from the raw background of 20 {ct}/s can be achieved through an Angular Resolution Measure (ARM) cut applied to the data. \citep{Zoglauer2006}. Note that applying an ARM cut also reduces the number of source counts. We account for this effect by scaling the flare counts according to the corresponding background reduction factor. Specifically, we estimate that for a factor of 20 reduction in the background rate (our baseline scenario), the source counts are reduced by a factor of $\sim 3$, while for a background reduction factor of 2, the corresponding source reduction factor is $\sim 1.2$.

Table \ref{tab:flare_mdp_summary} summarizes the number of flares for which we estimate $\mathrm{MDP}_{99\%}<50\%$ and $\mathrm{MDP}_{99\%}<20\%$ for combinations of background rate, X-ray spectral index variation ($\gamma_1$), and factor $\eta$. Under baseline background {and X-ray spectral} conditions, {no flares appear} as observable by COSI with $\mathrm{MDP}_{99\%}<50\%$.
%{only 1 flare appears as observable by COSI with $\mathrm{MDP}_{99\%}<50\%$}. 
{Nevertheless, under favorable (yet reasonably expected) spectral conditions ---namely, for a softer X-ray spectrum--- we estimate a substantially larger number of detectable flares with $\mathrm{MDP}_{99\%}<50\%$. For the baseline background rate of 1~{ct}/s, we find {31} flares for $\eta = 0.3$, increasing to 40 flares for $\eta = 0.5$, originating from 6 sources. The 6 sources for which we identify flares with achievable measurement of $\mathrm{MDP}_{99\%}<50\%$ are 4FGL~J0539.9-2839 (PKS~0537-286), 4FGL~J1129.8-1447 (PKS~1127-14), 4FGL~J1229.0+0202 (3C~273), 4FGL~J1256.1-0547 (3C~279), 4FGL~J2151.8-3027 (PKS~2149-306) and 4FGL~J2253.9+1609 (3C~454.3). Unsurprisingly, these 6 blazars correspond to the FSRQ subclass.} This is expected since FSRQs are not only typically brighter and more variable than BL Lacs \citep[see e.g.][]{hovatta2014,rajput2020}, but also have the peak of their high-energy SED hump at lower energies \citep[see][]{fossati1998,ghisellini2017}, closer to COSI's energy band. 

\begin{figure}
    \centering
    \includegraphics[height=6.5cm]{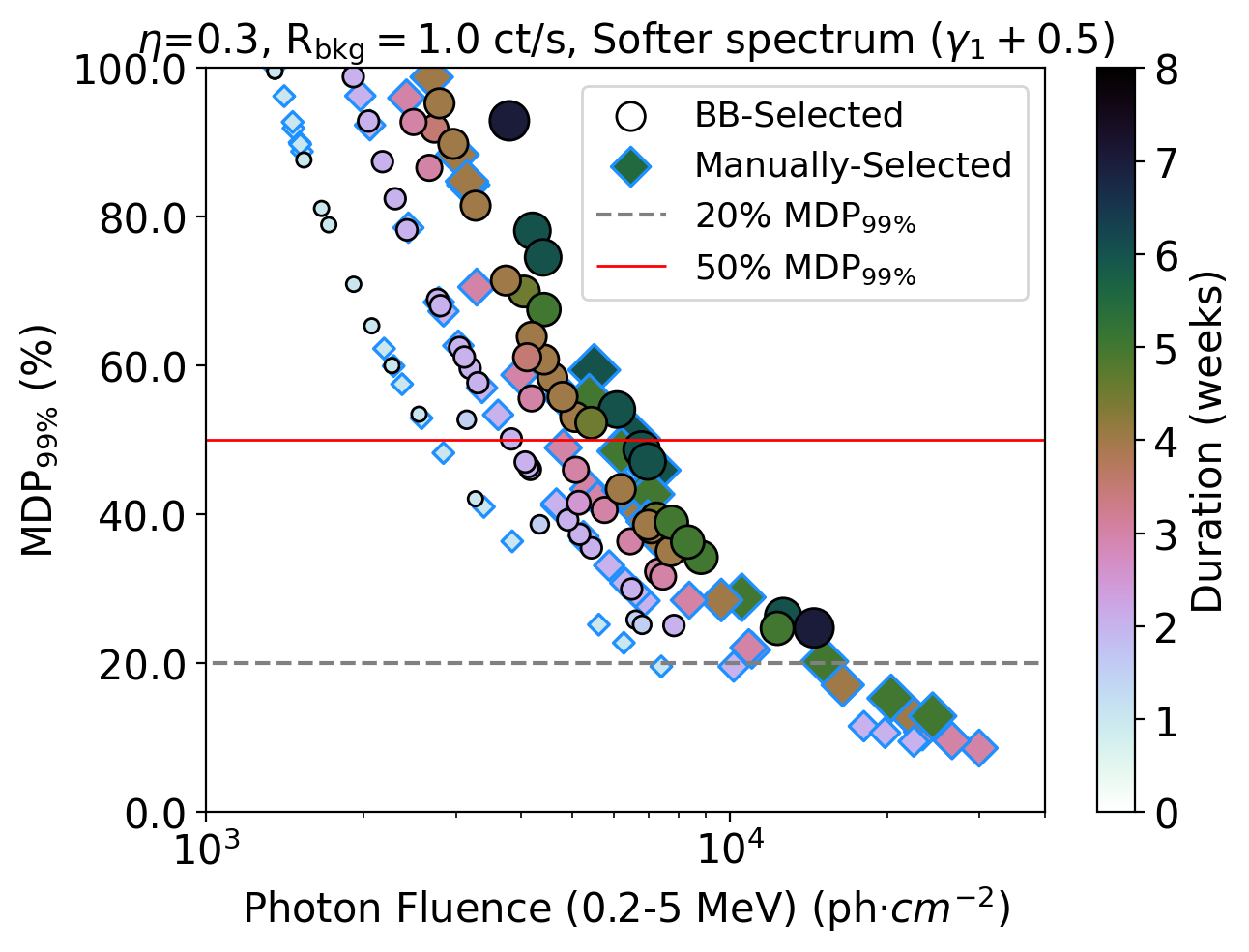}~~~~
     %\rule{6.5cm}{4cm} ~~~
    \includegraphics[height=6.5cm]{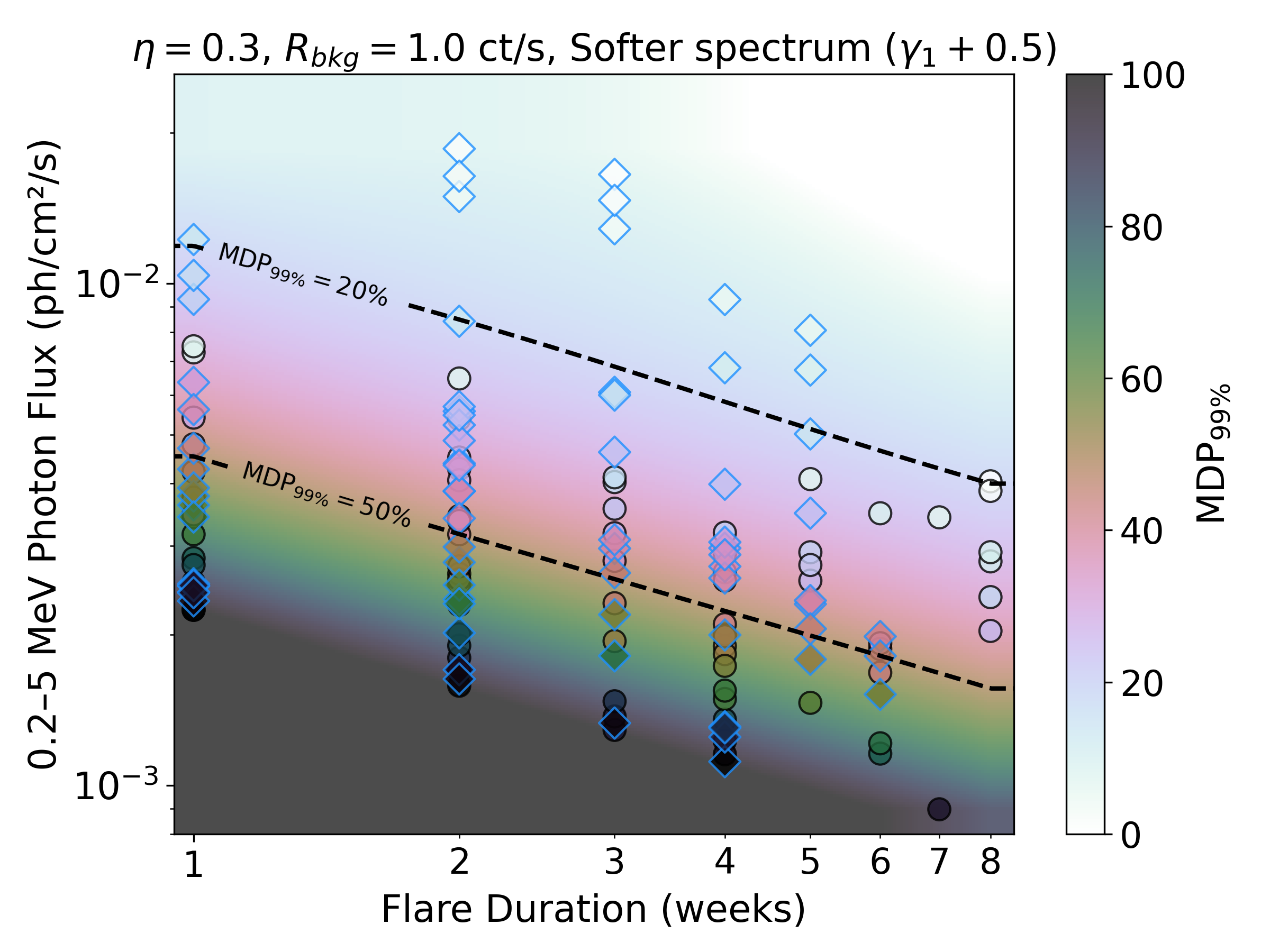}
    \caption{\textit{Left:} MDP$_{99\%}$ as a function of the photon fluence for every flare showing MDP$_{99\%}$ values below 100\%. The color scale and sizes of the markers represent the duration of the flare, with bigger markers representing longer flares and vice versa.  \textit{Right:} MDP$_{99\%}$ map for a grid of photon flux (0.2-5 MeV), assuming a background rate of 1 {ct}/s, $\eta=0.3$ and softer X-ray spectrum. {In both plots: circles mark flares detected by the automated flare detection method (BB-selected), while diamonds mark flares that were manually selected from the brightest parts of longer ($>$8 weeks) duration flares.} }
    \label{fig:pol_res}
\end{figure}

\begin{table}
\centering
\caption{Number of flares in our LCR selected sample with a duration $<$8 weeks identified to provide an MDP$_{99\%}$ in the COSI band below 50\% and 20\%. We report the results for combinations of $\eta$ and $\gamma_1$ and background rates, giving the number of detected flares and the corresponding number of sources in the format flares from sources. {We report the numbers separately for the BB-selection method and the manual selection method.}}
\begin{tabular}{c|c|c|c|c|c|c}
\hline
\multirow{2}{*}{Bkg rate $\mathrm{[{ct}/s]}$} & \multirow{2}{*}{$\gamma_1$} & \multirow{2}{*}{$\eta$} & \multicolumn{2}{c|}{BB-Selection} & \multicolumn{2}{c}{Manual Selection} \\
 & & & MDP$_{99\%}<50\%$ & MDP$_{99\%}<20\%$ & MDP$_{99\%}<50\%$ & MDP$_{99\%}<20\%$ \\
\hline
\multirow{6}{*}{20 (Raw Background)} & \multirow{3}{*}{Baseline} & 10\% & 0& 0&  &  \\
 & & 30\% & 0& 0& 5 from 2 sources& 0\\
 & & 50\% & 0& 0&  &  \\
\cline{2-7}
 & \multirow{3}{*}{+0.5 Softer} & 10\% & 0& 0&  &  \\
 & & 30\% & 0& 0& 28 from 9 sources& 8 from 3 sources\\
 & & 50\% & 0& 0&  &  \\
\hline
\multirow{6}{*}{10 (Pessimistic)} & \multirow{3}{*}{Baseline} & 10\% & 0& 0&  &  \\
 & & 30\% & 0& 0& 6 from 2 sources& 0\\
 & & 50\% & 0& 0&  &  \\
\cline{2-7}
 & \multirow{3}{*}{+0.5 Softer} & 10\% & 0& 0&  &  \\
 & & 30\% & 0& 0& 32 from 10 sources& 9 from 3 sources\\
 & & 50\% & 0& 0&  &  \\
\hline
\multirow{6}{*}{1 (Baseline)} & \multirow{3}{*}{Baseline} & 10\% & 0& 0&  &  \\
 & & 30\% & 0& 0& 9 from 4 sources& 0\\
 & & 50\% & 0& 0&  &  \\
\cline{2-7}
 & \multirow{3}{*}{+0.5 Softer} & 10\% & 16 from 3 sources& 0&  & \\
 & & 30\% & 31 from 6 sources& 0& 45 from 12 sources& 12 from 5 sources\\
 & & 50\% & 40 from 6 sources& 0&  &  \\
\hline
\end{tabular}
\label{tab:flare_mdp_summary}
\end{table}

As noted above, {so far we have restricted} our analysis to flares {identified by our BB analysis} with total durations shorter than 8~weeks. Motivated by the characteristic strong variability of polarization observed during blazar flaring episodes \citep{2018MNRAS.474.1296B}{, we adopt this duration cut as a loose upper limit for interesting flare duration for the present analysis. Previous studies have shown that blazars exhibiting more stable polarization in the optical band are predominantly BL~Lac objects, which, according to our results, are less favorable targets for COSI polarization detectability compared to FSRQs, whose polarization properties tend to be more variable \citep[][]{smith1996, angelakis2016, otero2023}. However, we note that, in some instances, longer-duration flares show several peaks or structures which could be further investigated and could potentially represent an important sample of flares populating the $\mathrm{MDP}_{99\%}<50\%$ or even $\mathrm{MDP}_{99\%}<20\%$ region. Therefore, we have inspected each individual $>$8 weeks flare for the brightest sources, and manually selected shorter periods of $\leq$6 weeks displaying a substantial flux increase. We detail this manual selection procedure in Appendix~\ref{long_flare_analysis}.} 

We illustrate the results for the baseline background rate (1 {ct}/s), $\eta$ = 0.3 and {softer X-ray} spectral conditions in Fig.~\ref{fig:pol_res} {with circle markers. Diamond markers represent flares that were manually selected from from the brightest parts of longer duration flares (duration $>$8 weeks.)}. On the left panel we show the $\mathrm{MDP}_{99\%}$ as a function of the flare fluence, $\mathcal{F}$ (ph/cm$^{2}$), defined as the product of the photon flux in ph/cm$^2$/s times the flare duration in seconds. On the right panel, we show the $0.2-5$ MeV flux as a function of the flare duration in weeks (for all flares $<$8 week long).

{The manual selection identifies several additional sources exhibiting flares with $\mathrm{MDP}{99\%}<50\%$, depending on the assumed background and spectral conditions, thereby improving the results reported in Table~\ref{tab:source_list}. For this subset, several sources remain promising even under pessimistic background scenarios of 10~ct/s, and two of them (4FGL~J1229.0+0202, 3C~273; and 4FGL~J2253.9+1609, 3C~454.3) show flares with $\mathrm{MDP}{99\%}<50\%$ even at a raw background rate of 20~ct/s. These findings are expected, as these events correspond to some of the brightest flares observed by the LAT, produced by the most luminous $\gamma$-ray blazars (see, e.g., Fig.~\ref{fig:lighcurve}). Overall, these results highlight the importance of optimizing the temporal selection around the brightest phases of flares in COSI analyses, which can be performed once the data is available by searching for the minimum $\mathrm{MDP}_{99\%}$. Incorporating this manual refinement alongside the initial BB analysis provides a more realistic assessment of  polarization detection prospects.}

{In our baseline background and spectral case, we count a total of 9 flares from 4 sources for which we obtain $\mathrm{MDP}_{99\%}<50\%$. For the most optimistic case of a favorable softer spectrum, the numbers increase to 45 additional flares from 12 sources, which corresponds to $\sim$6 flares expected during COSI's two-year prime mission.} 

{In Table~\ref{tab:source_list}, we list the 12 sources hosting flares with $\mathrm{MDP}_{99\%}<50\%$ (either identified by the BB analysis or by the manual selection) in the baseline scenario with $R_{\rm bkg} = 1$~{ct}/s, $\eta = 0.5$. As expected, this sample overlaps almost entirely with the blazars from which the LAT has detected the brightest $\gamma$-ray flares over its 17 years of operation. The 6 additional sources with promising flares identified with the manual selection are  {4FGL J0841.3+7053 (4C +71.07)}, 4FGL J1224.9+2122 (4C +21.35, PKS 1222+216), 4FGL J1332.0-0509 (PKS 1329-049), 4FGL J1512.8-0906 (PKS 1510-089), 4FGL J2202.7+4216 (BL Lacertae) and 4FGL J2232.6+1143 (CTA 102). Again, as expected, this sample is dominated by blazars of FSRQ type, with 4FGL J2202.7+4216 (BL Lacertae) being the only BL Lac of the list.}

\begin{table*}
\caption{List of sources, in alphabetical order, for which we identified flares with durations $<$8 weeks, {either with the BB method (BB-selected) or from the sample of bright long flares for which we have analyzed shorter time intervals (manually-selected),} and an achievable $\mathrm{MDP}_{99\%}<50\%$ by COSI under the {baseline} scenario of $\eta = 0.5$, $R_{\rm{bkg}}=1$~{ct}/s and a softer X-ray spectrum. Coordinates, associations and source classes correspond to those reported in the 4FGL-DR4 catalog.}
\label{tab:source_list}
\begin{center}
\hspace{-2cm}
\begin{tabular}{ccccccccc}
\hline
\multirow{2}{*}{} & \multirow{2}{*}{4FGL-DR4 name} & R.A. & Dec. & \multirow{2}{*}{Association name} & \multirow{2}{*}{Source class} & $\mathrm{MDP}_{99\%}<20\%$  &  $\mathrm{MDP}_{99\%}<20\%$ \\
 & & [deg] & [deg] &  &  & [BB-selected] & [Manually-selected] \\ \hline
1 & 4FGL J0539.9-2839 & 84.99 & -28.65 & PKS 0537-286 & FSRQ & No & No \\ %\hline
2 & 4FGL J0841.3+7053 & 130.34 & 70.88 & 4C +71.07  & FSRQ & {--} & Yes \\ %\hline
3 & 4FGL J1129.8-1447 & 172.46 & -14.79 & PKS 1127-14 & FSRQ & No & Yes \\ %\hline
4 & 4FGL J1224.9+2122 & 186.22 & 21.38 & 4C +21.35 & FSRQ & {--} & No \\ %\hline
5 & 4FGL J1229.0+0202 & 187.26 & 2.04 & 3C 273 & FSRQ & {No} & Yes \\ %\hline
6 & 4FGL J1256.1-0547 & 194.04 & -5.78 & 3C 279 & FSRQ & No & No \\ %\hline
7 & 4FGL J1332.0-0509 & 203.02 & -5.16 & PKS 1329-049 & FSRQ & -- & No \\ %\hline
8 & 4FGL J1512.8-0906 & 228.21 & -9.10 & PKS 1510-089 & FSRQ & -- & {No} \\ %\hline
9 & 4FGL J2151.8-3027 & 327.96 & -30.46 & PKS 2149-306 & FSRQ & {No} & {No} \\ %\hline
10 & 4FGL J2202.7+4216 & 330.69 & 42.28 & BL Lacertae & BL Lac & -- & No \\ %\hline
11 & 4FGL J2232.6+1143 & 338.15 & 11.73 & CTA 102 & FSRQ & -- & Yes \\ %\hline
12 & 4FGL J2253.9+1609 & 343.49 & 16.15 & 3C 454.3 & FSRQ & {No} & Yes \\ \hline
\end{tabular}
\end{center}
\end{table*}

\begin{figure}
     \centering
     \includegraphics[height=4.5cm]{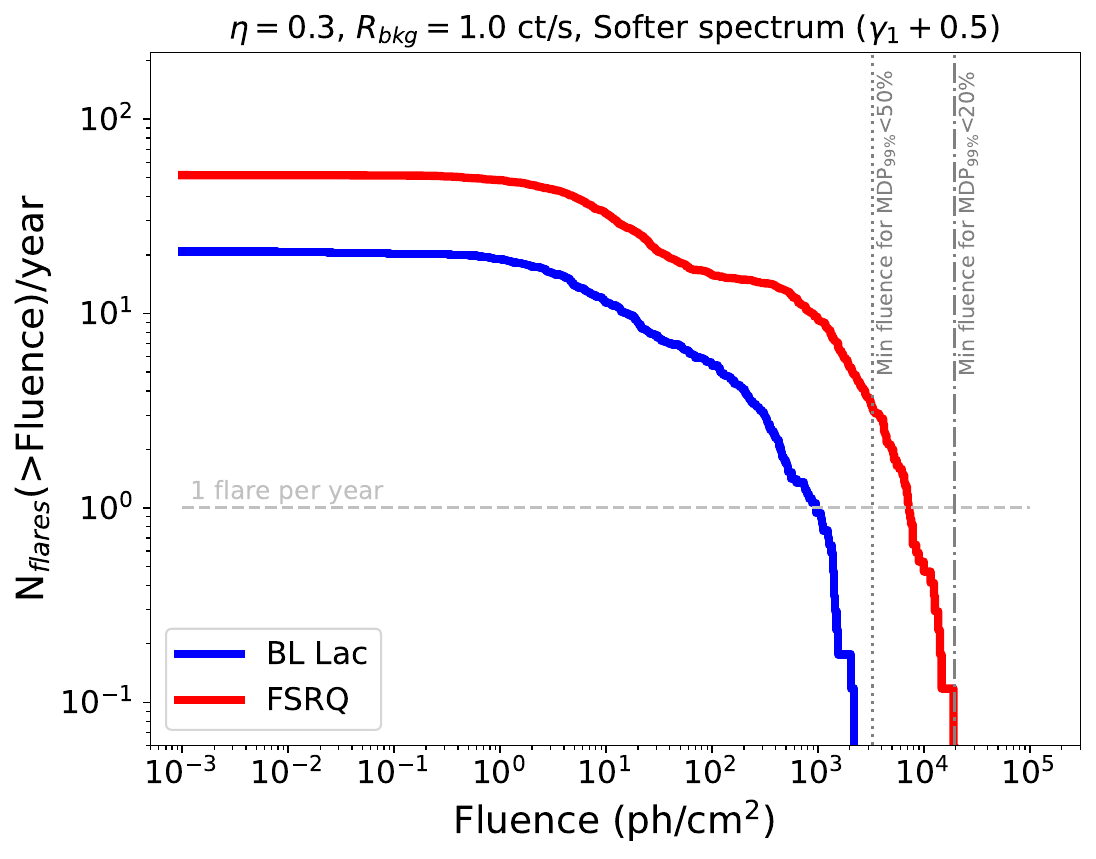}
     \includegraphics[height=4.5cm]{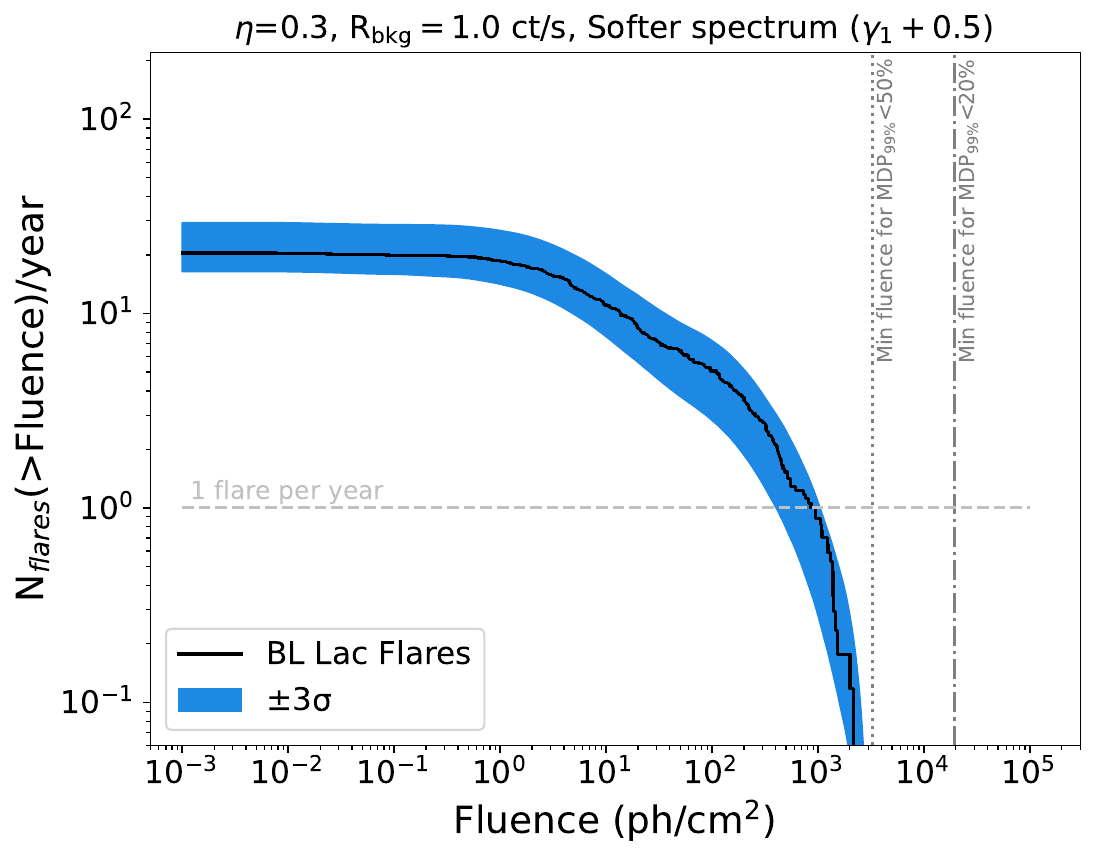}~~~
     \includegraphics[height=4.5cm]{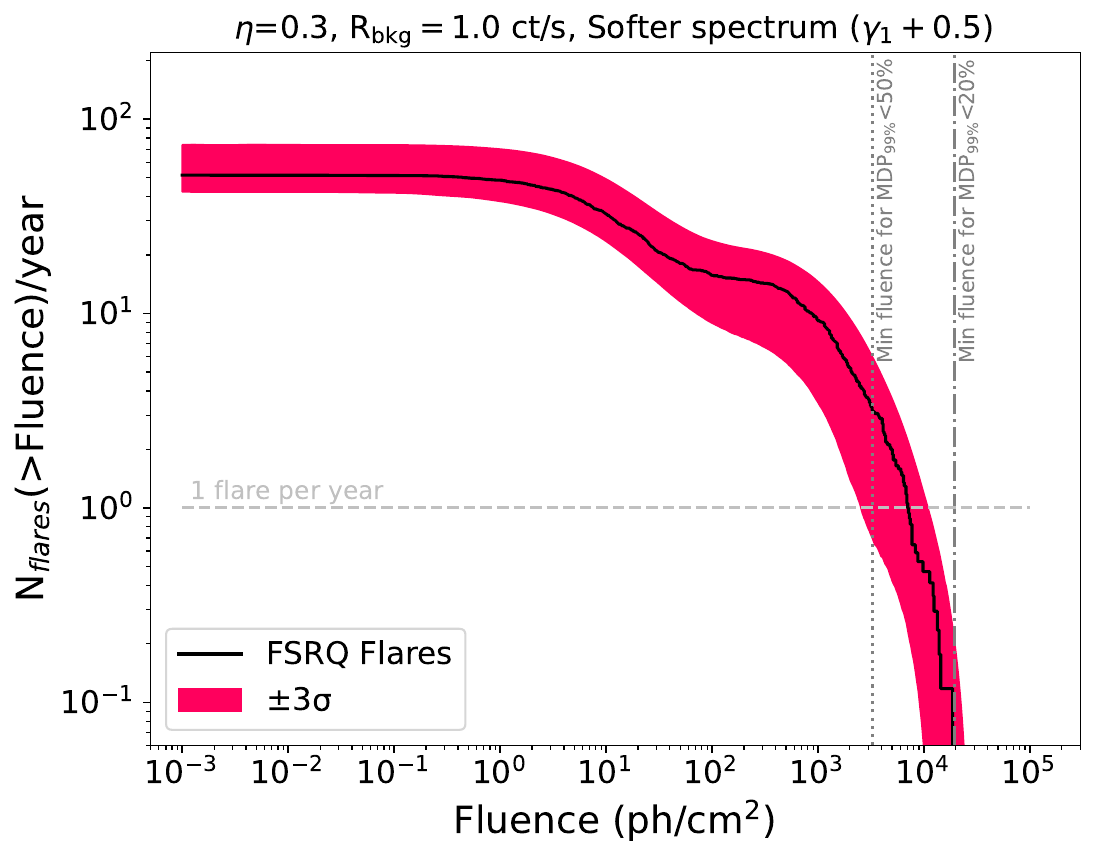}
     \caption{\textit{Left:} Blazars' flares fluence cumulative distributions for BL Lacs and FSRQs in the sample of flares with a duration $<$8 weeks obtained by setting $\eta=0.3$, $R_{\rm{bkg}}=1$ {ct}/s {and considering the case of a softer X-ray spectrum}. The horizontal dashed line corresponds to 1 flare per year. The dotted and dashed-dotted vertical lines correspond to the minimum fluences for which we achieve an $\mathrm{MDP}_{99\%}$ value of 50\% and 20\%, respectively, for $\eta=0.3$, $R_{\rm{bkg}}=1$ {ct}/s {and a softer X-ray spectrum}. \textit{Middle, Right:} Blazars' flare fluence distribution setting $\eta=0.3$, $R_{\rm{bkg}}=1$ ct/s {and considering the case of a softer X-ray spectrum}, and comparison with the expected distributions obtained from artificially simulated light curves generated with the method from \cite{emmanoulopoulos2013}, for BL Lac and FSRQs objects, respectively. The black solid lines correspond to the fluence distributions of the real data for each type of source and the colored contours are the $3\sigma$ confidence bands derived from the simulations. The horizontal dashed line indicates a frequency of 1 flare per year. The vertical dotted and dashed-dotted lines highlight the minimum fluences at which we estimate an $\mathrm{MDP}_{99\%}$ below 50\% and 20\% for a certain flare, respectively.}
     \label{fig:flare_fluence_simulations}
 \end{figure}

As discussed above, the $\mathrm{MDP}_{99\%}$ achieved for each flare depends on the total number of photons collected from the source, which is jointly determined by the flare brightness and its duration. Consequently, a longer but fainter flare can yield the same $\mathrm{MDP}_{99\%}$ as a shorter, brighter one. In this sense, requiring $\mathrm{MDP}_{99\%}$ to fall below a given threshold is equivalent to requiring that the total fluence of a flare, defined as the time-integrated photon flux, exceeds a corresponding threshold.
In our {baseline scenario} ($R_{\rm bkg} = 1$~{ct}/s, $\eta = 0.3$) {and considering a softer} spectrum, the minimum fluence for flares with $\mathrm{MDP}_{99\%}<50\%$ is approximately {$3.3 \times 10^{3}$~ph/cm$^{2}$}. As shown in the left panel of Fig.~\ref{fig:flare_fluence_simulations}, this implies that during COSI’s two-year prime mission we can expect {$\sim$5--6} flares from FSRQs for which COSI should be able to measure polarization below 50\% at 99\% confidence level. {In order to achieve polarization measurements with $\mathrm{MDP}_{99\%}\sim20\%$, these flares would need to be comparable to some of the brightest flares ever detected by \textit{Fermi}-LAT.}
%We also expect a smaller number of events (a few) for which COSI could achieve polarization measurements with $\mathrm{MDP}_{99\%}\sim20\%$.

To better estimate the uncertainty in this expected statistics, we carried out simulations of a population of blazars mimicking the statistical properties of the light curves in our sample. These simulations are performed for each blazar individually, with the method described by \cite{emmanoulopoulos2013} and the \textsc{python} \texttt{DELCgen} package implemented by \cite{connolly2015}. This approach proposes simulations that use the same statistical properties of the real light curve, that is, the same probability density function (PDF). Therefore, the artificial data series derived from the simulations have a flux distribution and variability amplitude consistent with those of the real data. In addition, this method uses as input a specific power spectral density (PSD) for the simulation process. For this, we use a PSD represented by a power law function to reproduce the same source of dominant noise in the long-term evolution of the artificial light curve, fitting in each case the PSD slope of the simulated source. As observed by previous studies, blazars often show PSDs consistent with a spectral slope $1<\beta<2$ {for $\gamma$-ray light curves} \citep[see e.g.,][]{finke2014,tarnopolski2020}, interpreted as a combination of the classical red noise that is known to dominate blazar variability on long timescales, and white noise that becomes more dominant for fast variability features \citep{otero2024}. The values fitted for our simulations are overall consistent with previous studies and typically fall within this range.
We apply this methodology, producing 1000 light curves for each real light curve retrieved from the \textit{Fermi}-LAT LCR. The same methodology of flare identification based on the BB analysis, detailed in Sect.~\ref{sec:flare_extraction}, is applied to each artificial light curve, assuming $\eta = 0.3$ as for the results presented in Fig.~\ref{fig:flare_fluence_simulations}. This enables us to calculate and compare the fluence distribution of each iteration of simulated light curves with that derived from the real data. We take as reference the baseline average spectrum case, with a background rate of 1~{ct}/s, {and apply this comparison to the flares identified by the BB analysis with a duration $<$8 weeks}. Fig.~\ref{fig:flare_fluence_simulations} illustrates the cumulative distribution of flare fluences for the simulated populations compared to the real distributions. We observe that the simulation approach is validated by the fact that the real fluence distribution falls well within the variance of simulations. 

These confidence bands also allow us to have an estimation of the uncertainty on the number of expected flares with or above a given fluence for each blazar type {and for the case of flares with a duration strictly $<$8 weeks}. Considering the fluence thresholds {$\mathcal{F}_{\rm{min}}^{\rm{MDP_{99\%}<50\%}}=3.3 \times 10^3$~ph/cm$^{2}$ and $\mathcal{F}_{\rm{min}}^{\rm{MDP_{99\%}<20\%}}=1.9\times 10^4$~ph/cm$^{2}$}, we identify from FSRQs a total number of {$3.2^{+2.7}_{-2.5}$ flares per year with $\mathrm{MDP}_{99\%}<50$\%, and $0.04^{+0.2}_{-0.2}$} flares per year with $\mathrm{MDP}_{99\%}<20$\%. On the other hand, the number of expected flares from BL Lac objects assuming this baseline $\eta$ and $R_{\rm{bkg}}$ parameters is consistent with zero, in agreement with all promising sources {identified in this work} being FSRQs, as reported before.

%-------------------------------------------------------------------------------

\section{Summary and Conclusion}
\label{sec:conclusion}
We evaluated the detectability of polarized $\gamma$-ray emission from blazars with {COSI}. Using 17 years of \textit{Fermi}-LAT monitoring data, we extracted the flaring episodes from the light curves of a large sample of blazars. We modeled the spectral behavior of each flare using either LAT-only or joint BAT–LAT spectral fits, extrapolated to the MeV band, and evaluated the corresponding minimum detectable polarization ($\mathrm{MDP}_{99\%}$) for each event under a range of instrumental and astrophysical conditions.

Our analysis shows that, over the 17-year-long flare statistics, under baseline background conditions of 1 {ct}/s {and assuming average X-ray spectra}, {no flares} would be detectable with $\mathrm{MDP}_{99\%}<50\%$. 

{Assuming X-ray spectral softening during flares, this result significantly improves, with up to {40 flares from six} FSRQs showing achievable $\mathrm{MDP}_{99\%}$ values below 50\%.}
After accounting for the 17-year duration of the LAT dataset, we estimate that during COSI’s prime mission the number of polarization-detectable blazar flares for which COSI will reach under the assumption of {a softer X-ray} spectrum $\mathrm{MDP}_{99\%}<50\%$ would be on the order $\sim$5-6 events, likely from FSRQs. We provide a list of the most promising known blazars we could expect bright flares from during COSI prime mission as well as the characteristics of the flares that would result in $\mathrm{MDP}_{99\%}<50\%$. We also crosscheck the expected number of flares per year for which COSI will potentially achieve with light curve simulations based on the real LAT data. We obtain consistent numbers with respect to the estimations derived from the $\mathrm{MDP}_{99\%}$ calculation, with {$3.2^{+2.7}_{-2.5}$} flares per year reaching the minimum fluence for an $\mathrm{MDP}_{99\%}<50\%$, and {$0.04^{+0.2}_{-0.2}$} flares per year with $\mathrm{MDP}_{99\%}<20\%$ from FSRQs, and none from BL Lac objects. {These numbers significantly increase when considering shorter ($<6$ weeks) time intervals around brightest peaks in flares that the BB analysis identifies as a single flare longer than $>$8 weeks. Under these considerations, the sample of promising sources doubles (from 6 to 12), leading to the expectation of a total of $61$ flares with $\mathrm{MDP}_{99\%}<50\%$ and $\sim$10 flares with $\mathrm{MDP}_{99\%}<20\%$. For some of the brightest blazar flares ever detected by the LAT, an $\mathrm{MDP}_{99\%}<50\%$ is also achievable under more pessimistic background conditions. As expected, almost all these sources are FSRQ-type blazars.}

Our results emphasize that improvements in background rejection or targeting softer, high-variability blazars substantially enhance the mission’s discovery potential. We note nevertheless that these results are based on \textit{Fermi}-LAT currently known sources. Sources that have not been detected so far in the MeV band --- either FSRQs peaking at lower energies than those observed by the LAT \citep{lister2015} or the hypothesized ultra extremely high-peaked BL Lacs \citep[UEHBLs, see][]{Sciaccaluga2025} in which the synchrotron emission would extend up to MeV energies --- could potentially lead to more detections of significant polarization than those estimated here.

At present, systematic monitoring of blazars in the MeV range is virtually nonexistent, leaving the duty-cycle behavior of MeV blazars largely unexplored. With its capability for {daily}, all-sky observations in the 0.2–5~MeV range, COSI will open a new observational window, enabling the first direct measurements of MeV blazar duty cycles and variability patterns. We plan to carry out such a study with COSI, establishing the first statistical characterization of MeV blazar activity and its connection to jet emission physics.

Overall, our results demonstrate that COSI will open a new observational window onto the polarization of blazar $\gamma$-ray emission, providing a first direct probe of jet magnetic-field geometry and emission processes in the MeV regime. Although the detectability of blazar polarization is sensitive to background performance and source spectral shape, the results presented here establish a realistic baseline for the expected number of polarization detections during the prime mission. These findings lay the foundation for future population-level polarization studies of AGN jets and motivate coordinated multiwavelength campaigns to maximize the scientific return of COSI’s blazar observations.

\begin{acknowledgments}
{The authors thank the anonymous referee for a valuable review that helped improve the results of the paper.} M.N. and G.A.L. thank the LaSPACE Undergraduate Research Assistantship (LURA) Program, which supported this project under award number 80NSSC20M0110. S.G. acknowledges support by DLR grant 50OO2218. 
% Resources supporting this work were provided by National High Performance Computing (NHR) South-West at Johannes Gutenberg University Mainz.
The Compton Spectrometer and Imager is a NASA Explorer project led by the University of California, Berkeley with funding from NASA under contract 80GSFC21C0059. Resources supporting this work were provided by the NASA High-End Computing (HEC) Program through the NASA Advanced Supercomputing (NAS) Division at Ames Research Center.
J.O.-S. acknowledges founding from the Istituto Nazionale di Fisica Nucleare (INFN) Cap. U.1.01.01.01.009. S.C. and I.L. were funded by the European Union ERC-2022-STG - BOOTES - 101076343. Views and opinions expressed are however those of the author(s) only and do not necessarily reflect those of the European Union or the European Research Council Executive Agency. Neither the European Union nor the granting authority can be held responsible for them.
S.G. acknowledges support by DLR grant 50OO2218. Resources supporting this work were provided by National High Performance Computing (NHR) South-West at Johannes Gutenberg University Mainz.
\end{acknowledgments}

\facilities{\textit{Fermi}, COSI, \textit{Swift}}

\software{\texttt{Lightcurves} \citep{wagner2022}, \texttt{DELCgen} \citep{connolly2015}, \texttt{Cosipy} \citep{Martinez2023}, \texttt{MEGALib}\citep{Zoglauer2008}}

\appendix

\section{Sky-fraction--weighted effective area}
\label{app:aeff}

\begin{figure}[b]
    \centering
    \includegraphics[height=5cm]{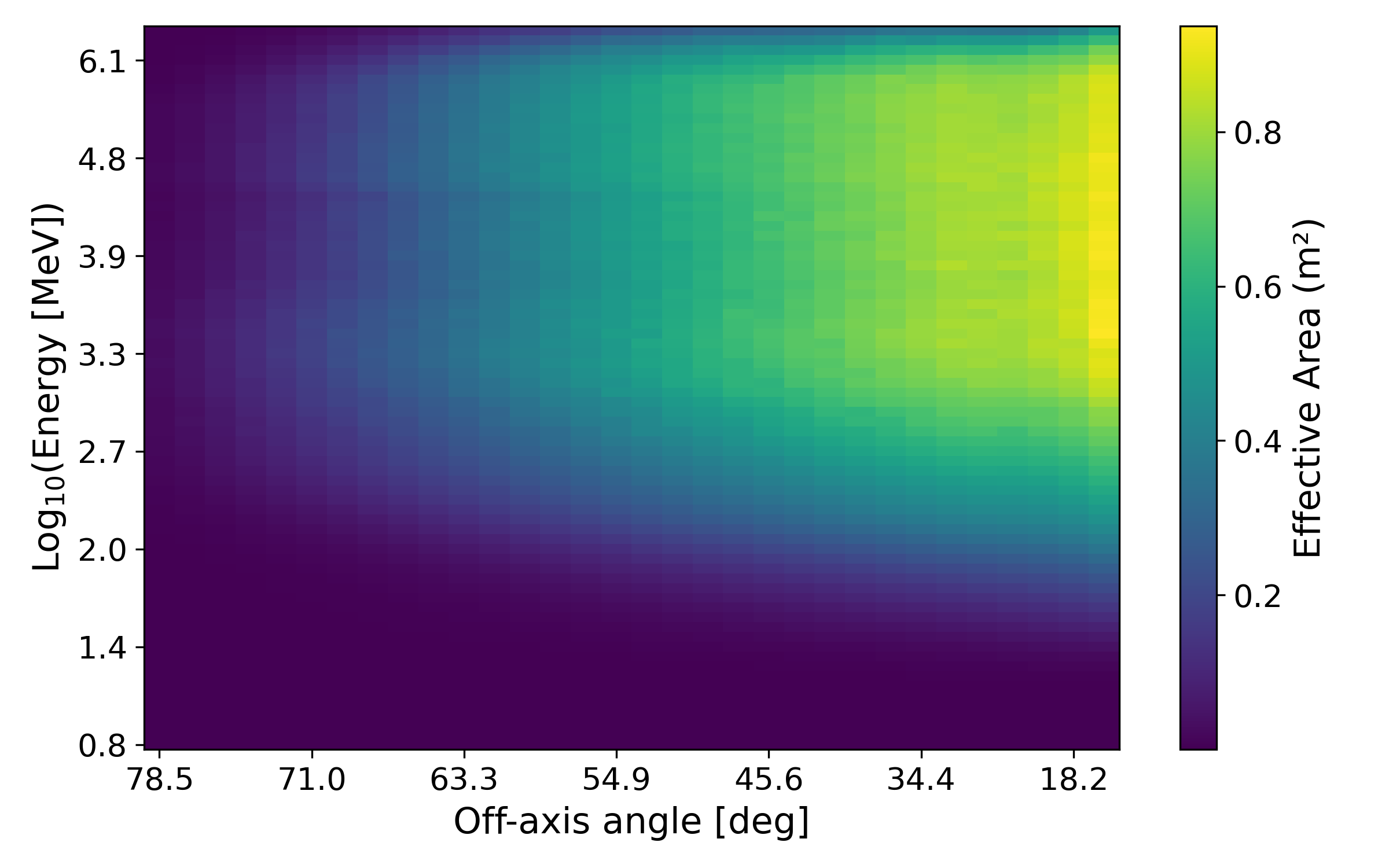} ~~~
    \includegraphics[height=5cm]{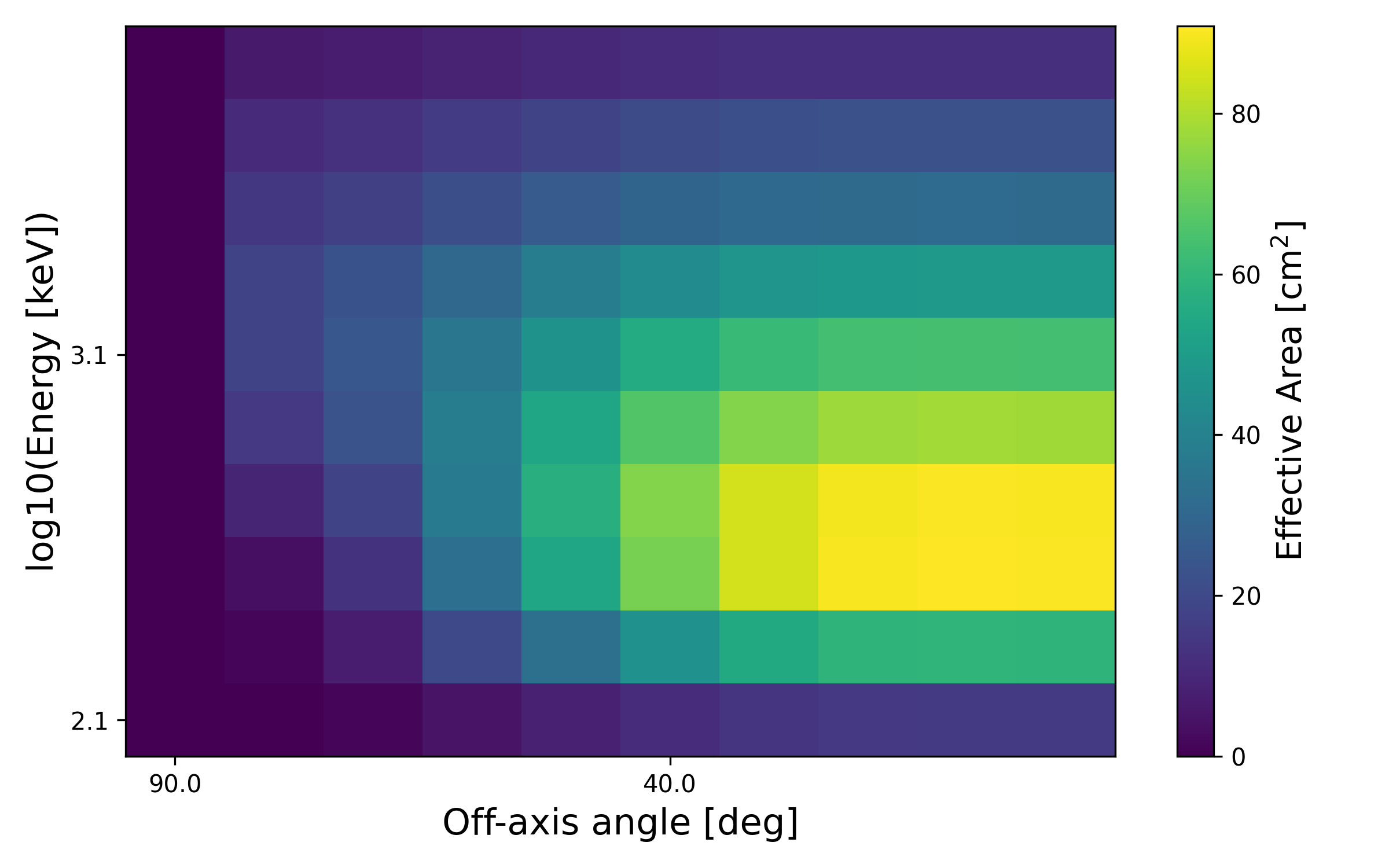}\\
    \includegraphics[height=5cm]{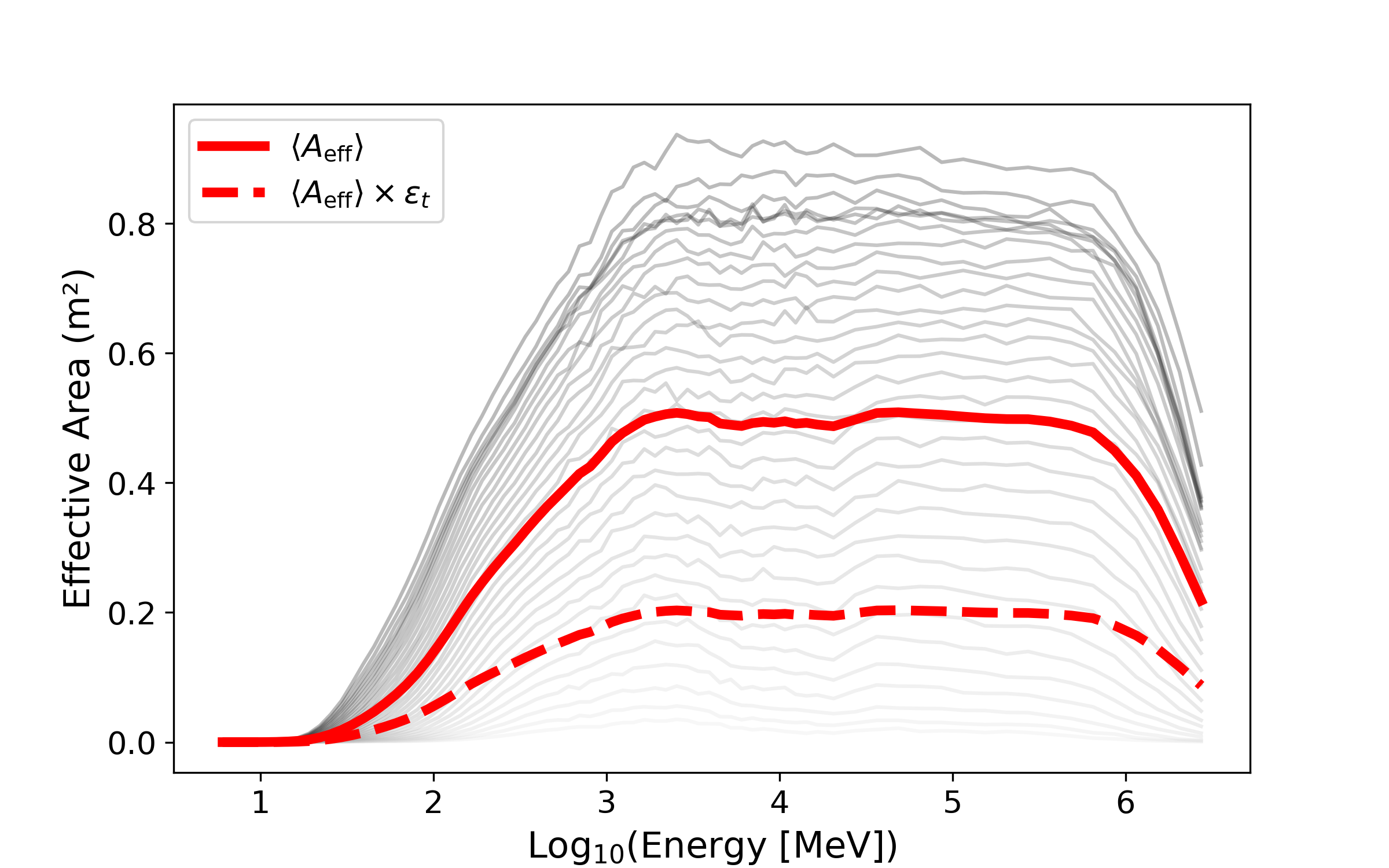} ~~~
    \includegraphics[height=5cm]{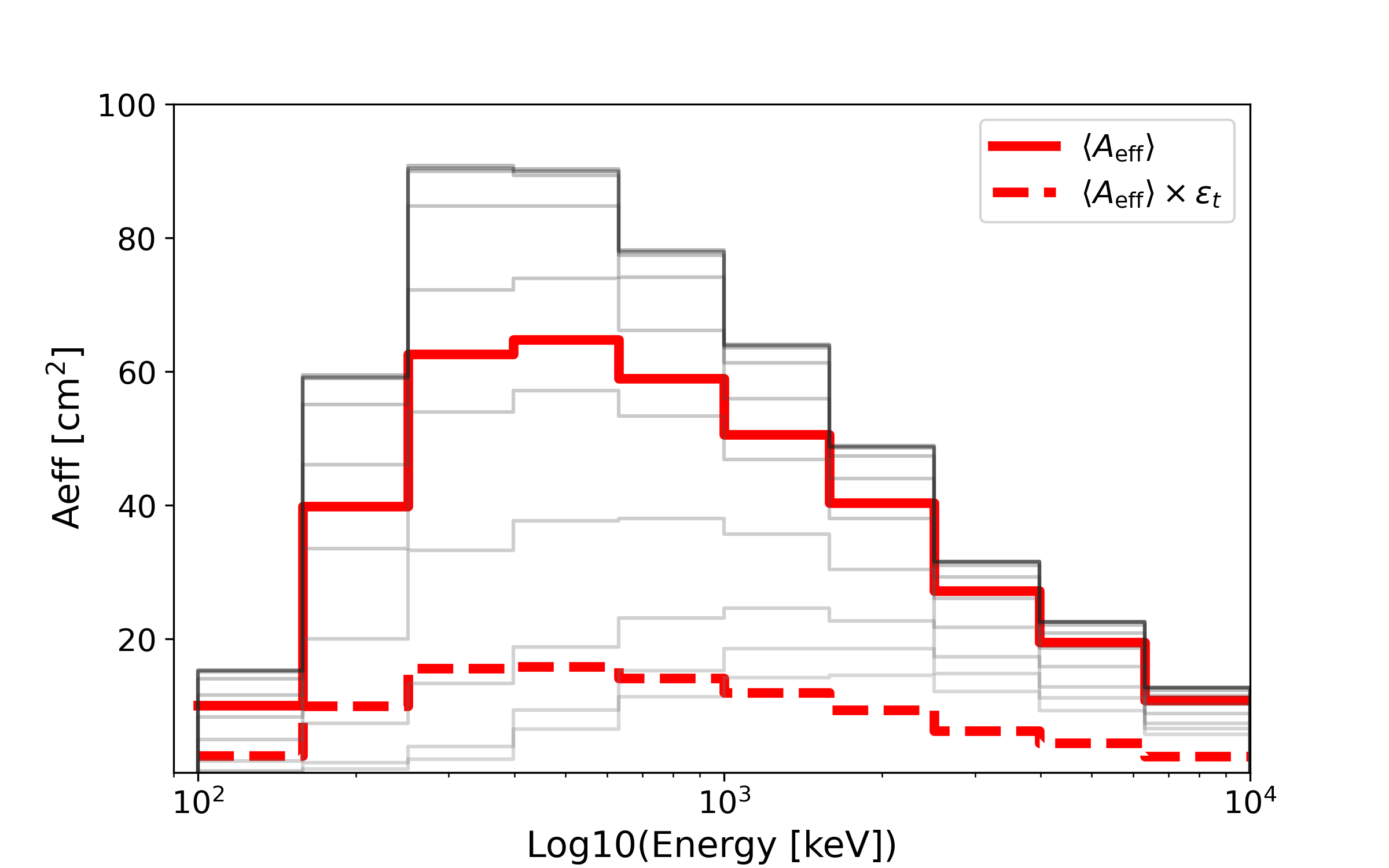}
    
    \caption{\textit{Top:} Effective area as a function of Energy and off-axis angle for \textit{Fermi}-LAT (left) and COSI (right). \textit{Bottom:} Effective area as a function of energy averaged over the off-axis angles considered in the analysis and multiplied by the effective exposure factor (in red), for \textit{Fermi}-LAT (left) and COSI (right). The gray lines show the effective areas at different off-axis angles as provided by the respective instruments (lighter {shades} to darker shades from more off-axis to on-axis).}
    \label{fig:aeffs}
\end{figure}

Fig.~\ref{fig:aeffs} shows the effective area as a function of energy and off-axis angle $\theta$, for both \textit{Fermi}-LAT (left) and COSI (right). Both instruments operate in survey mode, which implies that each given point source will be observed at different off-axis angles over time as the instruments orbit Earth. In particular, beyond a given off-axis angle, we can consider the effective area null, meaning that we would not consider the time intervals in which the source is at those angles.

We therefore compute the average effective area as 
\begin{equation}
% \boxed{
\langle A_{\mathrm{eff}}(E) \rangle_\theta =
\frac{
\displaystyle \sum_i
A_{\mathrm{eff},i}(E)
\left[\cos(\theta_{\mathrm{low},i}) - \cos(\theta_{\mathrm{high},i})\right]
}{
\displaystyle \sum_i
\left[\cos(\theta_{\mathrm{low},i}) - \cos(\theta_{\mathrm{high},i})\right]
},
% }
\end{equation}
where $i$ runs through off-axis angles and the formula is effectively a weighted average of the effective area over the off-axis angles accounting for the probability of a source being observed at each off-axis angle bin. 

\begin{figure}
    \centering
    \includegraphics[width=0.49\linewidth]{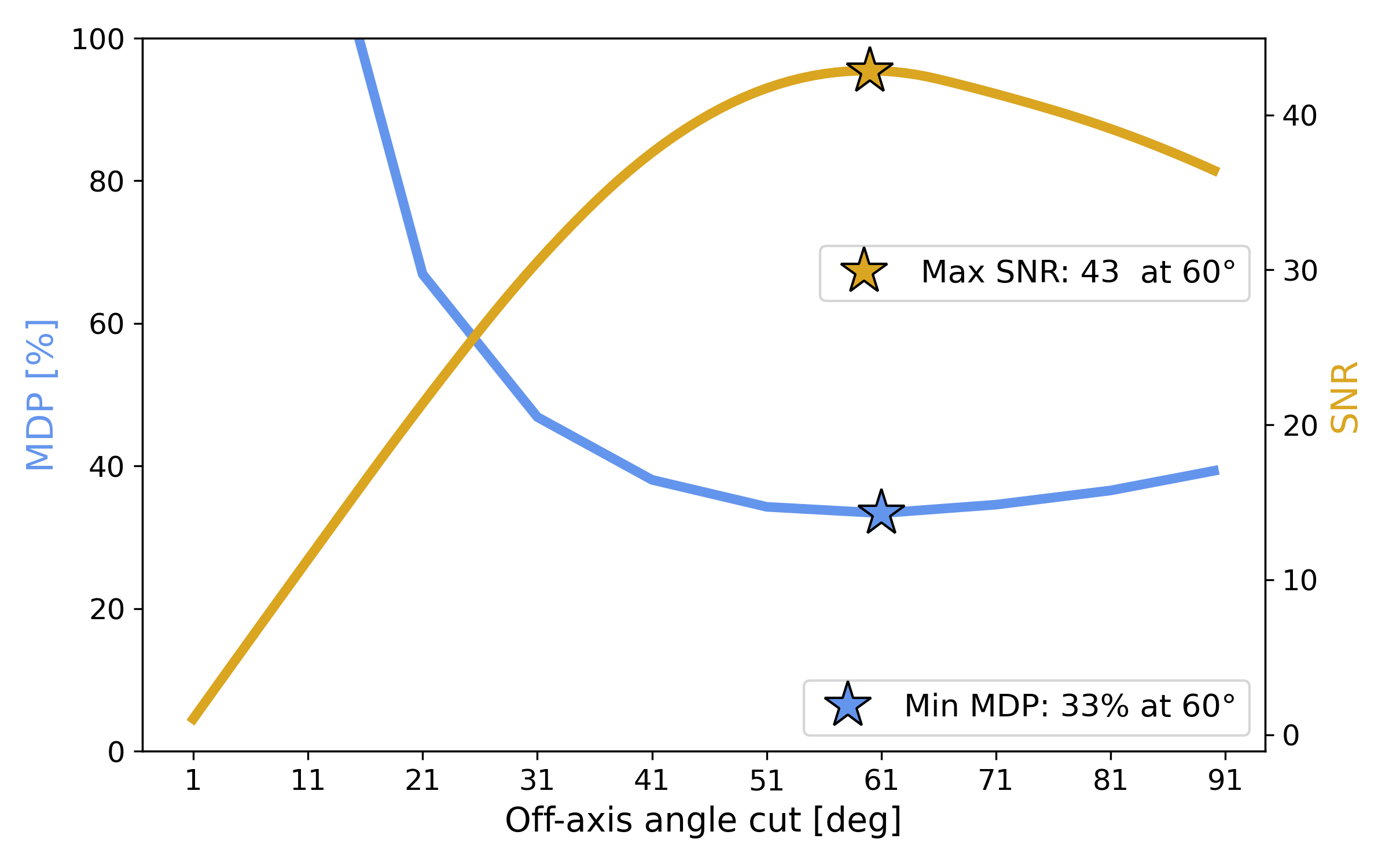}
    \includegraphics[width=0.49\linewidth]{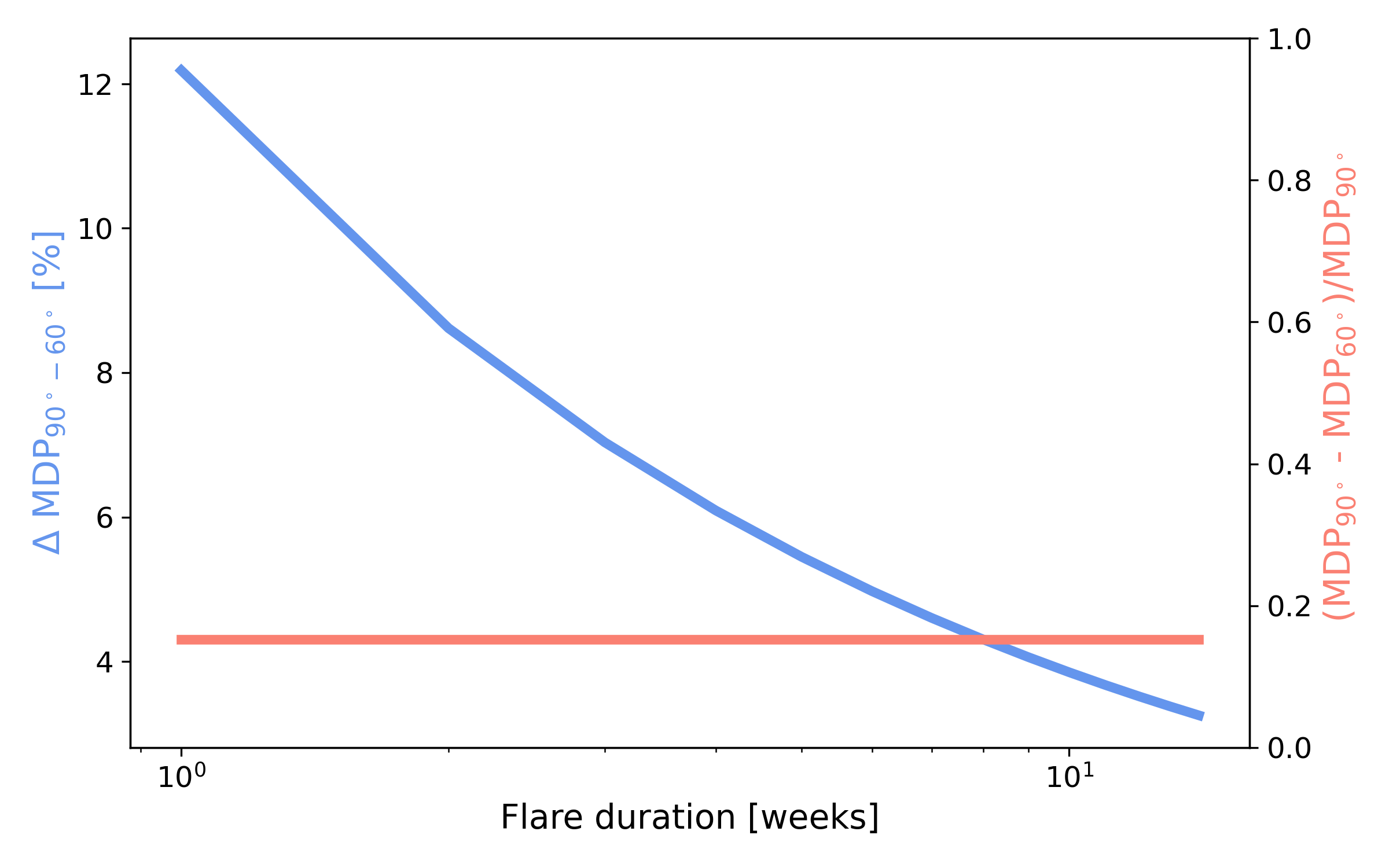}
    \caption{\textit{Left:} Dependence of the $\mathrm{MDP}_{99\%}$ (blue) and SNR (orange) on the maximum accepted off-axis angle. Both quantities reach their optimal values at $\sim60^\circ$, where the SNR {reaches a maximum value of 43} and the MDP reaches its minimum of 33\%. \textit{Right:} Comparison of the $\mathrm{MDP}_{99\%}$ improvement obtained by limiting the off-axis angle to $60^\circ$ instead of $90^\circ$. The absolute difference in $\mathrm{MDP}_{99\%}$ (blue, left axis) decreases with flare duration due to the $T^{-1/2}$ dependence of the $\mathrm{MDP}_{99\%}$, while the fractional gain (red, right axis) remains constant at $\approx15\%$, indicating that the benefit of the angular cut is independent of integration time once normalized (as expected for such long exposures).}
    \label{fig:offaxiscut}
\end{figure}

We compute the Average effective area in the visible hemisphere ($\Delta\theta=0^\circ-90^\circ$ for COSI, $\Delta\theta=0^\circ-78^\circ$ for the LAT), and we further account for the fraction of time the source would be in the other hemisphere with an effective exposure factor $\epsilon_t$:

\begin{equation}
    \mathcal{E}(E) = \langle A_{\mathrm{eff}}(E) \rangle_\theta \times \epsilon_t.
\end{equation}
$\epsilon_t$ is 0.25 for COSI, and 0.4 for \textit{Fermi}-LAT\footnote{This is given by the maximum off-axis angle provided in the CALDB for P8R3\_SOURCE IRFs (the same responses used to build the LCR).}. For COSI, the definitive off-axis angle cut for source analyses will be refined closer to launch. For this study, we assessed the optimal off-axis to be {60$^{\circ}$} by studying the variation of the signal-to-noise ratio (SNR) and the $\mathrm{MDP}_{99\%}$ for a simulated source with simple power-law spectrum and varying the off-axis cut, as illustrated in Fig.~\ref{fig:offaxiscut}.

\section{COSI Polarization response simulations}
\label{app:1}
\begin{figure}
    \centering
    \includegraphics[height=5cm]{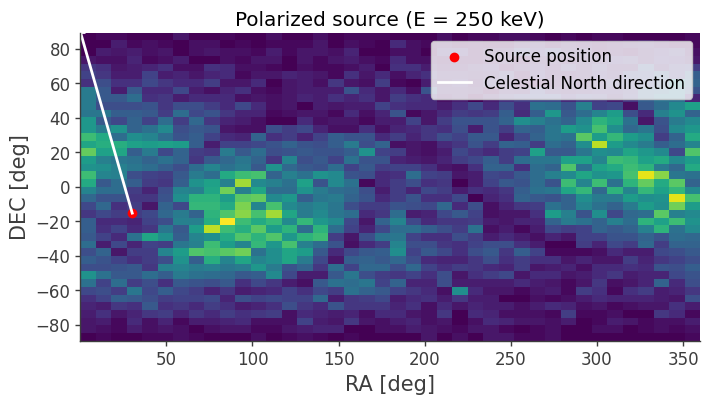}~~~~~~~~
    \includegraphics[height=5cm]{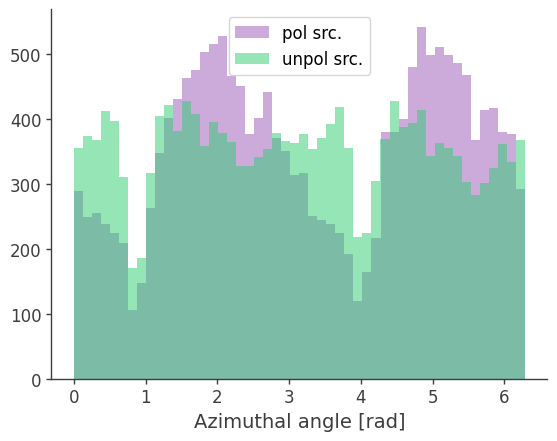} \\
    \includegraphics[height=5cm]{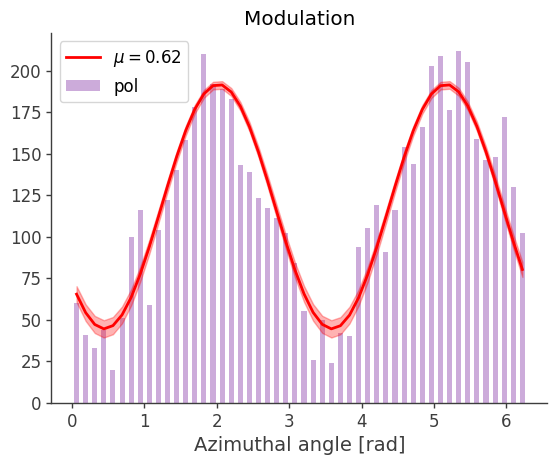}~~~~~~~~~
    \includegraphics[height=5cm]{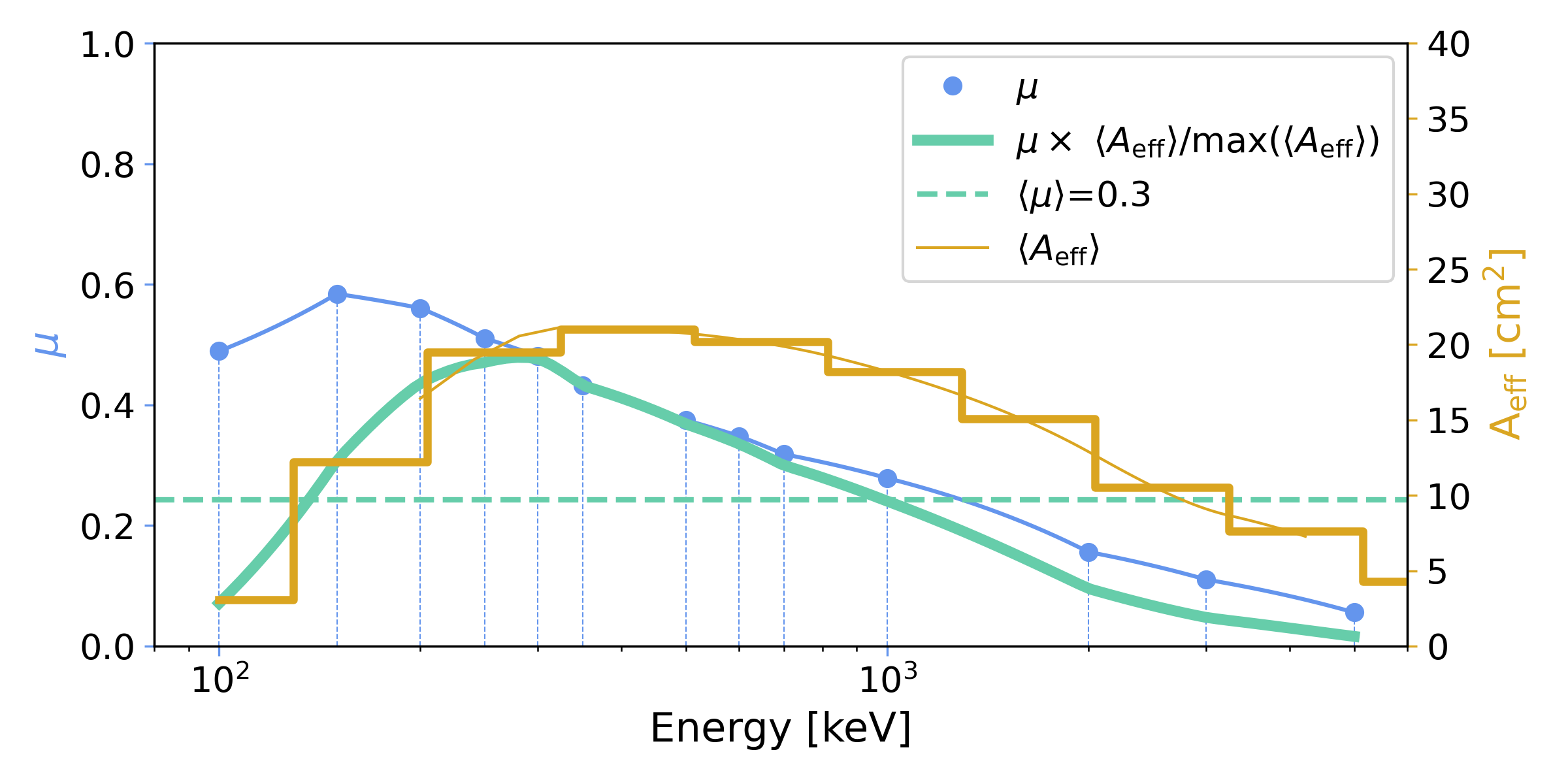}

    \caption{Simulation of a polarized source at 250 keV as seen by COSI SMEX. \textit{Top left:} Reconstructed sky map with the source location marked in red and the celestial north direction indicated and used as reference to define the azimuthal angle (increasing anti-clock-wise and ranging between 0 and 360 degrees). \textit{Top right:} Azimuthal scattering angle distributions for a polarized (purple) and unpolarized (green) source. \textit{Bottom left:} Modulation curve of the polarized source with best-fit sinusoidal model, yielding a modulation factor of $\mu=0.62$. \textit{Bottom right:} Energy dependence of the modulation factor and effective area. The modulation factor derived from monochromatic simulations (blue points with dashed vertical markers) and spline interpolation (blue curve) is shown on the left axis. The instrument's effective area is plotted on the right axis as both binned (step curve) and interpolated (orange line) values. The green curve shows the product $\mu \times A_{\rm eff}$, which serves as a proxy for the polarization sensitivity across energy.}
    \label{fig:mu}
\end{figure}

We studied the energy dependence of the polarization response of the COSI instrument, assuming the current design adopted in the DC3. The modulation factor, $\mu(E)$ is the response to 100\% polarized source of photons. To study the energy dependence we reconstruct the polarization degree of simulated 100\% polarized monochromatic sources $(E_{beam}$ = [100, 150, 200, 250, 300, 350, 500, 600, 700, 1000, 2000, 3000, 5000] keV). To estimate the polarization degree we look at the azimuthal angle distribution of the Compton-scattered photons interacting in the COSI detector and fit with a cosine square modulating function which is expected in case of detectable polarization. For this study we utilize existing COSIpy \citep{martinez-castellanos2023} routines. The fitting function is defined as 
$$R(x) = A + B*(cos(x + C)^2)$$
with A, B and C free parameters. The modulation factor is given by
$$ \mu = \frac{(R_{max}-R_{min})}{(R_{max}+R_{min})} $$
where $R_{min}$ and $R_{max}$ are the minimum and maximum value of the $R(x)$ evaluated at the best-fit values of the parameters. Because of the {color{red} asymmetry} (not a perfect sphere) of the COSI's instrument, we fit the polarized distribution subtracted (bin-per-bin of azimuthal angle) by an unpolarized distribution simulated with the same set up as the polarized case but with a polarization degree equal to 0. The modulation factor as a function of the energy is shown in Fig.~\ref{fig:mu} (bottom-right panel). Convolving by the COSI effective area and averaging in the $0.2-5$ MeV range, we obtain an average $<\mu> \sim 0.3$.

\section{Flare identification pipeline validation}
\label{app:2}
\begin{figure}
\centering
       \includegraphics[width = 0.75\linewidth]{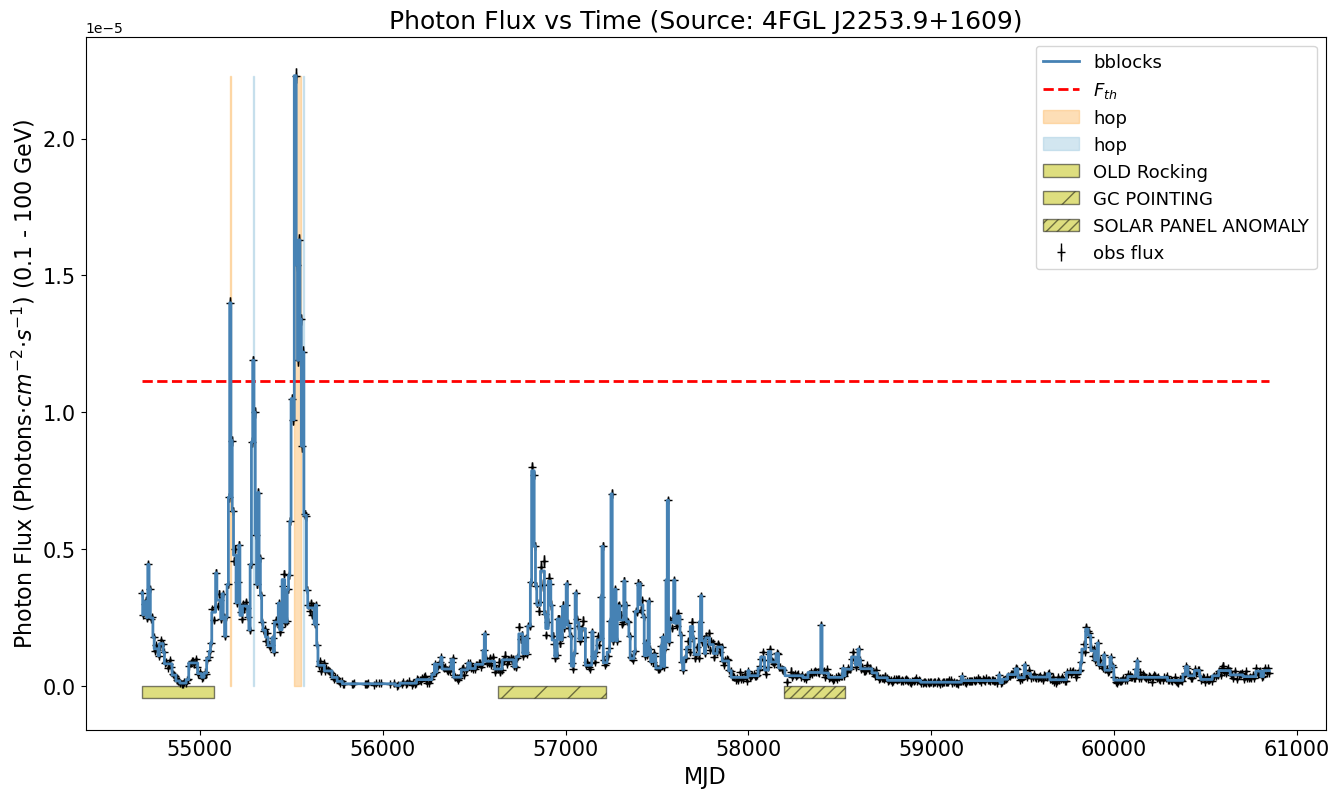} 
    \includegraphics[width = 0.75\linewidth]{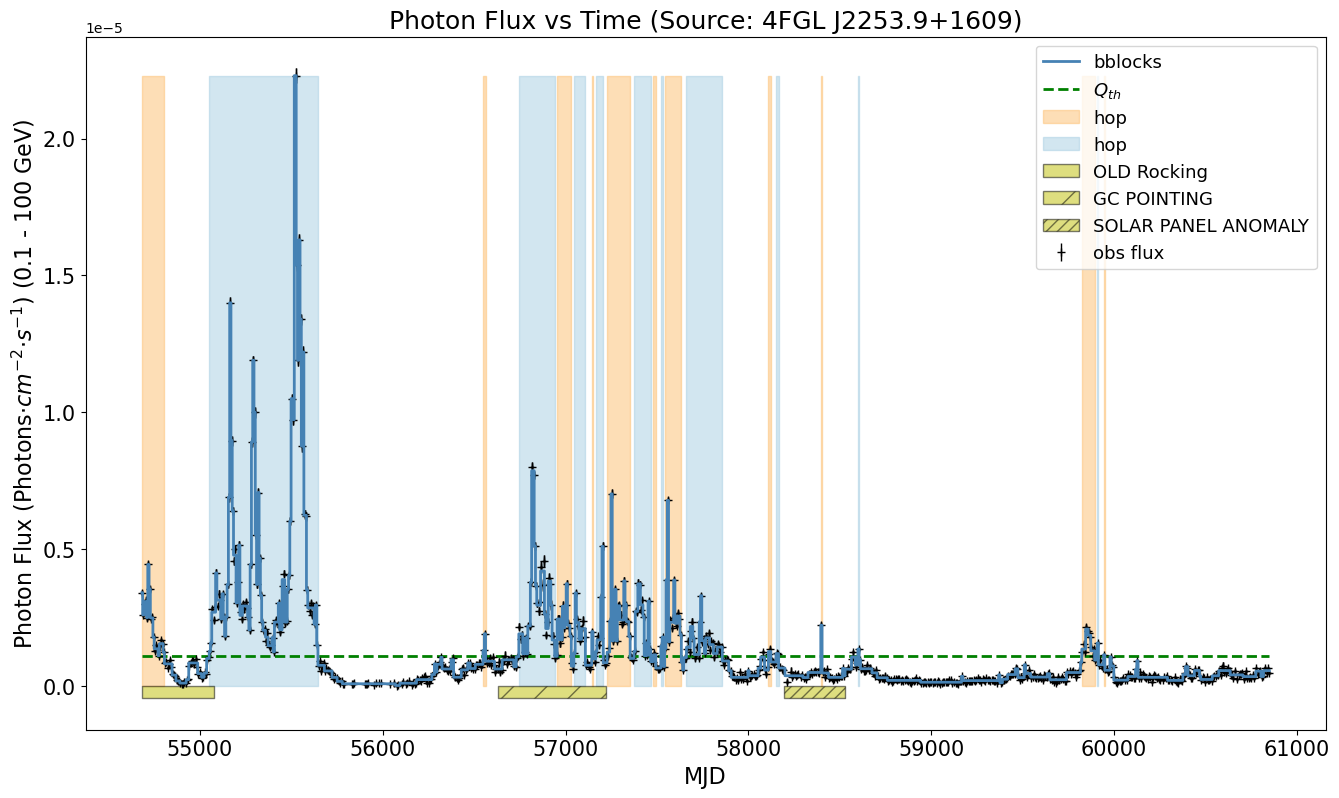}
    \caption{Photon flux light curve of 4FGL J2253+1609 using  $\eta=0.3$. The red dashed line is the initial value of $F_{th}$, and the green dashed line is the value of $Q_{th}$. The colored blocks are time periods that rise above and subsequently fall below the quiescent {threshold} level, $Q_{th}$.}
    \label{fig:threshold_illust}
\end{figure}

Our values for $\eta$ were chosen to be 0.1, 0.3, and 0.5. As shown in Eq.~(\ref{eq:flux_threshold}), a higher value of $\eta$ links to a higher $F_{th}$. To illustrate the process of calculating $F_{th}$ and $Q_{th}$, Fig.~\ref{fig:threshold_illust} illustrates the light curve and BB analysis of the source 4FGL 2253.9+1609 for the case of $\eta = 0.3$.

As stated before, the different colored shaded regions represent the detected flare periods. Larger $\eta$ values implies a higher $F_{th}$ and, consequently, because the non-flaring periods would be longer and averaging flux values with larger variance, a higher $Q_{th}$ is also expected.
% Here we can put the additional polarization tests.

In order to verify that our pipeline identifies flares correctly, we look at the statistics of flares examining the $\gamma$-ray duty cycle of the blazars in our sample, that is, the fraction of time that they spend in a flaring state, and compare with values in literature. For this test, we do not apply any cut in flare duration and consider the full statistics of identified flares to match more closely previous studies of blazar's duty cycle. We adopt the definition of flaring states described in Section~\ref{sec:flare_extraction}, and define the duty cycle as
\begin{equation}
    \bar{D}^{\rm src} = \frac{\sum_{\rm{i}}d^{\rm src}_{\rm i}}{\Delta t_{\rm obs}},
\end{equation}
where $i$ is the index running on the number of flares {identified} for the source $src$, $d^{\rm src}_{\rm i}$ is the duration of the flare $i$, and $\Delta t_{\rm obs}$ is the total observation time (17 years).
For this study, we focus on the case with $\eta = 0.3$, $R_{\rm{bkg}}$=1 {ct}/s as a benchmark case, and we perform this study separately on our sample of BL Lacs and FSRQs. The resulting flare sample comprises 595 blazars in total, including 205 BL~Lacs and 356 FSRQs.

On top of the standard duty cycle, which answers the question ``What fraction of mission time did this source spend in a flaring state?'', we are interested in ``What fraction of time did the source spend in its brighter flaring states?'', which is a better diagnostic to determine polarization detectability,
\begin{equation}
    \bar{D}^{\rm src}_{w} = \frac{\sum_{\rm{flares}}d^{\rm src}_{\rm i}}{\Delta t_{\rm obs}\sum_{\rm{i}}w_{\rm i}},
    \label{eq:weightedDC}
\end{equation}
where $w_i = F_{\rm flare, i}/{\rm max}(F_{\rm flares})$, $F_{\rm flare, i}$ being the average flux of the individual flare and ${\rm max}(F_{\rm flares})$ is the maximum flux value among the average flux values of the full blazar population\footnote{We tested alternative definitions of the flux weighting, normalizing the photon flux either globally across the full sample, separately for each blazar class, or individually per source. The resulting cumulative distributions remain nearly unchanged. The main reason for such a stability is the normalization for the weights on a source basis as indicated in Eq. (\ref{eq:weightedDC}).}.

\begin{figure}
    \centering
    \includegraphics[height=6.5cm]{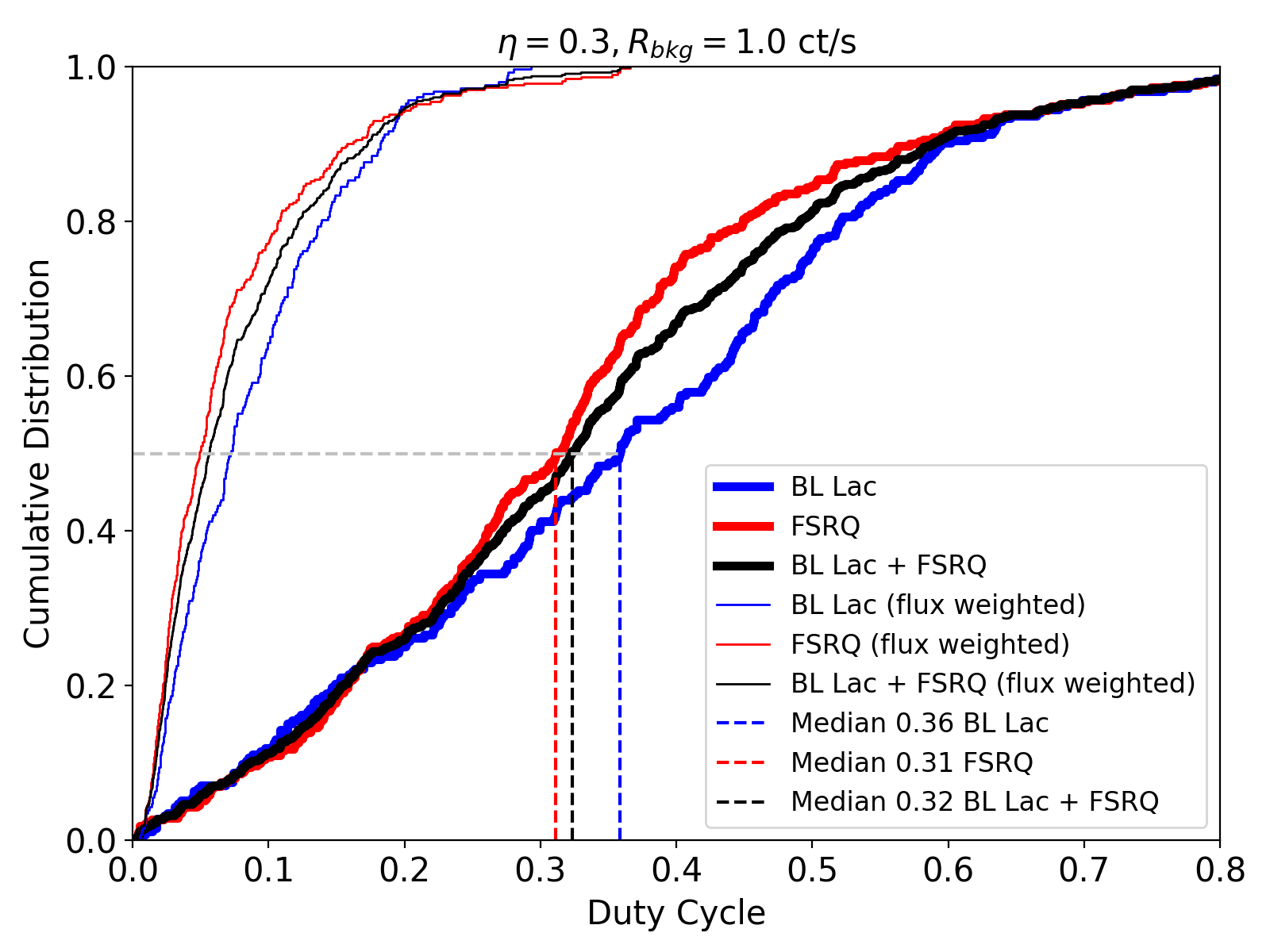}
    \caption{Cumulative distribution function of the duty cycles (thick lines) and flux-weighted duty cycle (thin lines) for BL Lacs (blue), FSRQs (red), and the combined population of BL Lacs and FSRQs (black) for the case with $\eta=0.3$, $R_{\rm{bkg}}=1$~{ct}/s. The medians of the distributions are also reported in the legend.}
    \label{fig:duty}
    \end{figure}

Fig. \ref{fig:duty} shows the cumulative distributions of the duty cycles, both unweighted (thick lines) and flux-weighted (thin lines) for BL Lacs (in blue) and FSRQs (in red). 

Performing a two-sample Kolmogorov-Smirnov \citep[or Anderson–Darling,][]{AndersonDarling1954} test on the unweighted duty cycle, we find a maximum deviation $D_{KS}$ of 0.182, {which allows us to reject the null hypothesis that the two samples are coming from the same population at} more than 5$\sigma$ ($D_{\alpha=10^{-7}}=0.14$). The median values of the two distributions are 0.36 (BL Lacs) and 0.31 (FSRQ), respectively, with FSRQs displaying on average smaller duty cycles than BL Lacs, indicating that the former are active for a smaller fraction of time. 

Applying the Anderson–Darling test to the \emph{flux–weighted} duty–cycle cumulative distributions yields a similar test statistic of 0.203. Emphasizing high–flux intervals thus shifts the distributions to lower duty cycles (as expected), with BL Lacs and FSRQs experiencing a similar contraction\footnote{We quantify the global leftward shift between the unweighted and weighted curves as
$\Delta D \equiv \bigl|\tilde{D}_{\mathrm{w}}-\tilde{D}\bigr|$, where
$\tilde{D}$ denotes the sample median. We find
$\langle \Delta D_{\rm BL~Lac}\rangle \simeq 0.29$ and
$\langle \Delta D_{\rm FSRQ}\rangle \simeq 0.26$.
The non-parametric Mann–Whitney U test \citep{MannWhitney1947} comparing the $\Delta D$ distributions returns $p=0.02$, indicating that the two populations shift left-ward similarly.}.
Overall, a larger number of bright flares are expected from FSRQ than from BL Lacs. Consistently with this picture, all flares in our sample with $\mathrm{MDP}_{99\%}<50\%$ are associated with FSRQs. The duty cycle values we find are in line with other studies of blazars duty cycle carried out with independent flare definitions \citep[see, e.g.,][]{Sacahui2021,yoshida2023}. It is important to note that the duty cycle estimation depends by definition on the selected blazar sample and the flare identification procedure, and therefore, it is not surprising that other studies with different assumptions for defining flaring states and different samples of blazars show different values.

\section{Filtering of misidentified flares}
\begin{figure}
\centering
    \includegraphics[width = \linewidth]{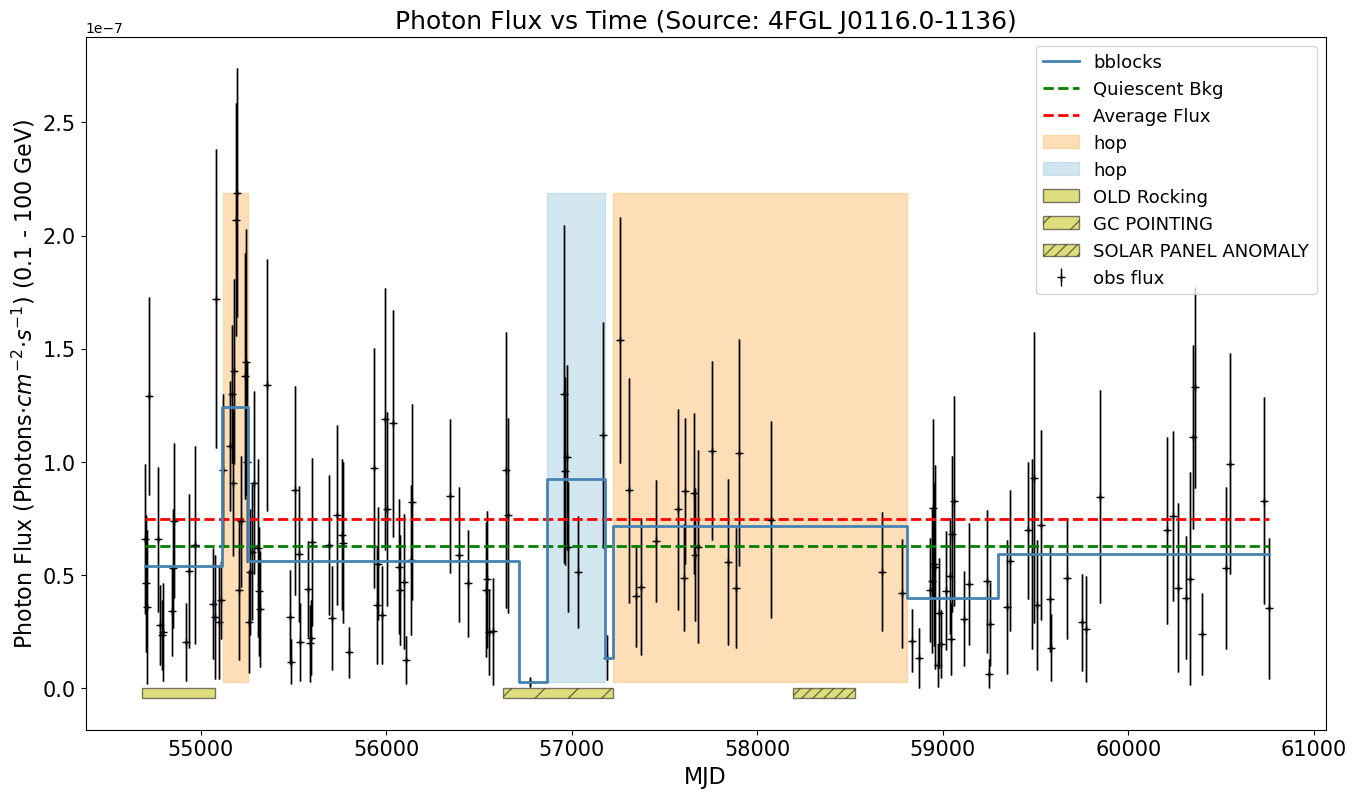} 
    \caption{Example of the light curve of blazar 4FGL J0116.0-1136 with fake flare identifications. The first \texttt{hop} corresponds to a real flare, while the second and third are clearly consistent with constant flux, but identified as flares by the BB algorithm due to the two flux points before them, with very low value and small uncertainties, most likely cause by fluctuations due to low statistics.}
    \label{fig:fake_flares}
\end{figure}

The flare identification relies on the public LAT light curves from the LCR. We have identified that, for some very faint sources, the BB analysis led a misidentification of significant variability in the light curves. This is due to very faint flux points with abnormally small uncertainties, introduced by fluctuations due to low statistics and/or short effective exposures. This causes the BB algorithm to evaluate the next flux point as a potential real variability, identifying periods consistent with constant emission as flares (see Fig.~\ref{fig:fake_flares}). As can be seen in the example below, the average flux of these cases is approximately the same as the quiescent {threshold} $Q_{th}$ defined for the BB analysis, allowing us to identify these ``fake'' flares. Hence, we have filtered out from our $\mathrm{MDP}_{99\%}$ analysis these cases.

\section{Promising candidates for $\mathrm{MDP}_{99\%}<50\%$}

In Table~\ref{tab:brightest_flares}, we compile all the sources showing flares with duration $<$8 weeks from which COSI could achieve an $\mathrm{MDP}_{99\%}<50\%$, within the {baseline background case} ($\eta = 0.3$, $R_{\rm{bkg}}=1$~{ct}/s) {and a softer X-ray spectrum} defined in our analysis. For each flare, we characterize its $\mathrm{MDP}_{99\%}$, average and peak flux in the COSI band, and duration in weeks. In Table~\ref{tab:flare_list}, we have compiled all of the flares from Fig.~\ref{fig:pol_res}, and listed their source, class, and associated flux values.

\begin{table}
\caption{List of the brightest flare from each source identified to produce flares below 50\% MDP$_{99\%}$ for COSI for our baseline {background} case ($\eta = 0.3$ , $R_{\rm{bkg}}=1$~{ct}/s) {and a softer X-ray spectrum} with a flare duration of $<$8 weeks within Fig.~\ref{fig:pol_res}. We report the associated average flare flux, peak flare flux, and flare duration within COSI's energy band (0.2-5 MeV). Each named source and respective classes correspond to those reported in the 4FGL-DR4 catalog.}
\begin{center}
\hspace{-1.7cm}
\begin{tabular}{cllcccc} 
\hline
\multirow{2}{*}{} & \multirow{2}{*}{4FGL-DR4 name} & \multirow{2}{*}{Class} & MDP$_{99\%}$& Average Flux& Peak Flare Flux & Duration
\\ 
&  &  & [\%]& [10$^{-3} \times$~ph/cm$^2$/s]& [10$^{-3} \times$~ph/cm$^2$/s]& [Weeks]\\ 
\hline
%1 & 4FGL J1031.6+6019 & FSRQ & 15.6 & 8.68 & 8.68 & 3.0 \\
1 & 4FGL J2253.9+1609 & FSRQ & 19.7 & 4.04 & 5.38 & 8.0 \\
2 & 4FGL J1229.0+0202 & FSRQ & 20.5 & 3.86 & 6.82 & 8.0 \\
%12 & 4FGL J0427.3-3900 & BCU & 28.3 & 8.22 & 8.22 & 1.0 \\
3 & 4FGL J1129.8-1447 & FSRQ & 29.9 & 5.37 & 7.69 & 2.0 \\
4 & 4FGL J0841.3+7053 & FSRQ & 38.5 & 2.03 & 3.20 & 8.0 \\
5 & 4FGL J2151.8-3027 & FSRQ & 38.5 & 2.89 & 3.05 & 4.0 \\
6 & 4FGL J0539.9-2839 & FSRQ & 43.3 & 2.56 & 3.83 & 4.0 \\
7 & 4FGL J1256.1-0547 & FSRQ & 48.6 & 1.87 & 4.04 & 6.0 \\
\hline
\end{tabular}
\label{tab:brightest_flares}
\end{center}
\end{table}

\begin{table}
\caption{List of BB-selected flares identified below 50\% MDP$_{99\%}$ for COSI for our {baseline background case} ($\eta = 0.3$, $R_{\rm{bkg}}=1$~{ct}/s) {and a softer X-ray spectrum} within Fig.~\ref{fig:pol_res}, with their associated MDP$_{99\%}$, average and peak flare flux, and flare duration within COSI's energy band (0.2-5 MeV). Each named source and respective classes correspond to those reported in the 4FGL-DR4 catalog. The flares organized in order of increasing MDP$_{99\%}$.}
\begin{center}
\hspace{-1.7cm}
\begin{tabular}{cllcccc} 
\hline
\multirow{2}{*}{} & \multirow{2}{*}{4FGL-DR4 name} & \multirow{2}{*}{Class} & MDP$_{99\%}$& Average Flux& Peak Flare Flux & Duration
\\ &  &  & [\%]& [10$^{-3} \times$~ph/cm$^2$/s]& [10$^{-3} \times$~ph/cm$^2$/s]& [Weeks]\\ 
\hline
%1 & 4FGL J1031.6+6019 & FSRQ & 15.6 & 8.68 & 8.68 & 3.0 \\
%1 & 4FGL J2253.9+1609 & FSRQ & 19.7 & 4.04 & 5.38 & 8.0 \\
%2 & 4FGL J1229.0+0202 & FSRQ & 20.5 & 3.86 & 6.82 & 8.0 \\
1 & 4FGL J1229.0+0202 & FSRQ & 24.6 & 4.08 & 5.52 & 5.0 \\
2 & 4FGL J1229.0+0202 & FSRQ & 24.7 & 3.42 & 4.43 & 7.0 \\
3 & 4FGL J1229.0+0202 & FSRQ & 25.0 & 6.47 & 6.47 & 2.0 \\
4 & 4FGL J1229.0+0202 & FSRQ & 25.1 & 7.50 & 7.50 & 2.0 \\
5 & 4FGL J1229.0+0202 & FSRQ & 25.8 & 7.29 & 7.29 & 2.0 \\
6 & 4FGL J2253.9+1609 & FSRQ & 26.3 & 3.49 & 7.27 & 6.0 \\
%9 & 4FGL J1229.0+0202 & FSRQ & 27.0 & 2.92 & 3.71 & 8.0 \\
%10 & 4FGL J1229.0+0202 & FSRQ & 28.1 & 2.80 & 4.17 & 8.0 \\
%12 & 4FGL J0427.3-3900 & BCU & 28.3 & 8.22 & 8.22 & 1.0 \\
%11 & 4FGL J1031.6+6019 & FSRQ & 29.3 & 2.80 & 2.80 & 8.0 \\
7 & 4FGL J1129.8-1447 & FSRQ & 29.9 & 5.37 & 7.69 & 2.0 \\
8 & 4FGL J1229.0+0202 & FSRQ & 31.6 & 4.11 & 4.70 & 3.0 \\
9 & 4FGL J1229.0+0202 & FSRQ & 32.2 & 4.02 & 4.96 & 3.0 \\
%15 & 4FGL J1229.0+0202 & FSRQ & 33.0 & 2.37 & 2.70 & 8.0 \\
10 & 4FGL J1229.0+0202 & FSRQ & 34.2 & 2.92 & 3.89 & 5.0 \\
11 & 4FGL J1229.0+0202 & FSRQ & 35.0 & 3.19 & 3.84 & 4.0 \\
12 & 4FGL J1229.0+0202 & FSRQ & 35.4 & 4.50 & 5.31 & 2.0 \\
13 & 4FGL J1129.8-1447 & FSRQ & 36.2 & 2.75 & 4.48 & 5.0 \\
14 & 4FGL J1229.0+0202 & FSRQ & 36.3 & 3.56 & 4.43 & 3.0 \\
15 & 4FGL J1229.0+0202 & FSRQ & 37.3 & 4.27 & 4.44 & 2.0 \\
16 & 4FGL J1229.0+0202 & FSRQ & 38.0 & 2.93 & 3.73 & 4.0 \\
%23 & 4FGL J0841.3+7053 & FSRQ & 38.5 & 2.03 & 3.20 & 8.0 \\
17 & 4FGL J2151.8-3027 & FSRQ & 38.5 & 2.89 & 3.05 & 4.0 \\
18 & 4FGL J1229.0+0202 & FSRQ & 38.6 & 4.78 & 4.78 & 2.0 \\
19 & 4FGL J1129.8-1447 & FSRQ & 38.8 & 2.56 & 3.89 & 5.0 \\
20 & 4FGL J1229.0+0202 & FSRQ & 39.2 & 4.06 & 5.06 & 2.0 \\
21 & 4FGL J1229.0+0202 & FSRQ & 39.4 & 2.66 & 3.18 & 5.0 \\
22 & 4FGL J1229.0+0202 & FSRQ & 40.5 & 3.18 & 3.43 & 3.0 \\
23 & 4FGL J2151.8-3027 & FSRQ & 41.5 & 3.41 & 3.78 & 3.0 \\
24 & 4FGL J1229.0+0202 & FSRQ & 42.0 & 5.41 & 5.41 & 1.0 \\
25 & 4FGL J0539.9-2839 & FSRQ & 43.3 & 2.56 & 3.83 & 4.0 \\
26 & 4FGL J1129.8-1447 & FSRQ & 45.9 & 2.80 & 3.31 & 3.0 \\
27 & 4FGL J1229.0+0202 & FSRQ & 45.9 & 3.44 & 3.82 & 2.0 \\
28 & 4FGL J1229.0+0202 & FSRQ & 46.3 & 3.41 & 3.63 & 2.0 \\
29 & 4FGL J2151.8-3027 & FSRQ & 46.9 & 3.36 & 3.59 & 2.0 \\
30 & 4FGL J1129.8-1447 & FSRQ & 47.0 & 1.92 & 2.63 & 6.0 \\
31 & 4FGL J1256.1-0547 & FSRQ & 48.6 & 1.87 & 4.04 & 6.0 \\
\hline
\label{tab:flare_list}
\end{tabular} 
\end{center}
\end{table}

\section{Manual selection of long-flare sub-intervals}\label{long_flare_analysis}
{In order to evaluate a realistic data analysis in which we can select a specific time interval to be analyzed from an entire blazar light curve, we perform a manual selection of the brightest peaks in sub-intervals shorter than 6 weeks inspecting $>$8 weeks flares identified with the BB analysis.} 

{First we identify all bright sources with flares $>$8 weeks and a maximum flux $\geq$10$^{-6}$~cm$^{-2}$~s$^{-1}$. We identify 25 sources above this maximum flux with at least one of these long-lasting flares: 4FGL~J0221.1+3556, 4FGL~J0319.8+4130, 4FGL J0403.9-3605, 4FGL~J0442.6-0017, 4FGL~J0539.9-2839, 4FGL~J0841.3+7053, 4FGL~J0904.9-5734, 4FGL~J1129.8-1447, 4FGL~J1153.4+4931, 4FGL~J1159.5+2914, 4FGL~J1224.9+2122, 4FGL~J1229.0+0202, 4FGL~J1256.1-0547, 4FGL~J1332.0-0509, 4FGL~J1512.8-0906, 4FGL~J1522.1+3144, 4FGL~J1642.9+3948, 4FGL~J1740.5+5211, 4FGL~J1800.6+7828, 4FGL~J1833.6-2103, 4FGL~J2056.2-4714, 4FGL~J2151.8-3027, 4FGL~J2202.7+4216, 4FGL~J2232.6+1143, and 4FGL~J2253.9+1609. }

{For each of these long flares, we run the HOP identification procedure on the BB lightcurve raising the flux threshold (manually set on a case-by-case basis) to isolate the brightest peaks in the sub-structure of each long-flare. We add to the sample of BB-identified flares only the peaks with $MDP<100\%$ and duration $\leq$6 weeks. We refer to the flare sample obtained from this procedure as ``manually-selected'' flares to distinguish them from the BB-selected ones. An example of this selection is shown below in Fig.~\ref{fig:manual_selection} for the source 4FGL~J2253.9+1609.}

\begin{figure}
\centering
    \includegraphics[width = \linewidth]{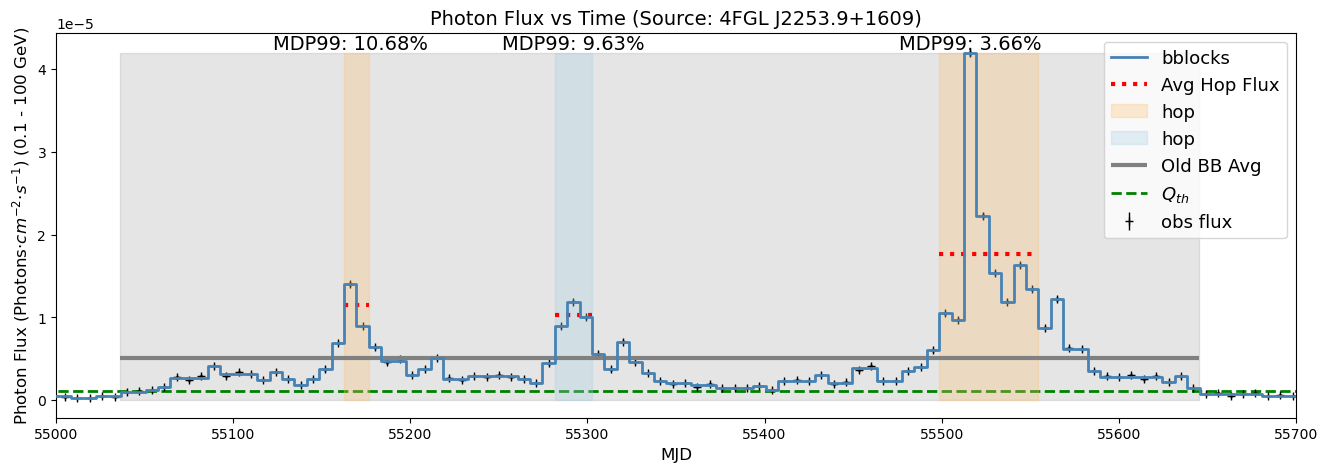} 
    \caption{{Example of a Long-flare (gray shaded region) identified in the Fermi-LAT light curve of source 4FGL J2253.9+1609. Manually-selected time windows with bright peaks are highlighted in orange and blue shades. Dotted red lines represent the average flux for those shorter time windows (all higher than the global average of the BB-identified flare, shown with a solid gray line). Above each manually selected flare we report the MDP$_{99\%}$ value after extrapolation to the COSI band. Dashed, horizontal green line shows the BB-calculated quiescent background for the source.}}
    \label{fig:manual_selection}
\end{figure}

\bibliography{sample7}{}
\bibliographystyle{aasjournalv7}

\end{document}